\documentclass[
    11pt,
    letterpaper,
    reprint,
    nofootinbib,
    notitlepage,
    superscriptaddress,
    eqsecnum,
    aps,
    prx,
    nobalancelastpage
]{revtex4-2}

\usepackage{amsmath,amssymb,amsfonts} %
\usepackage{empheq}
\usepackage{physics}

\usepackage{amsthm}	%

\newtheorem{theorem}{Theorem}
\newtheorem{lemma}{Lemma}

\usepackage{subdepth} 	%

\usepackage{bm}	%
\renewcommand{\mathbf}{\bm}
\usepackage{dsfont}	%
\renewcommand{\mathbb}{\mathds}	%

\usepackage[svgnames,dvipsnames]{xcolor}
\usepackage{graphicx}
\definecolor{NewBlue}{rgb}{0.1, 0.1, 0.7}
\definecolor{NewRed}{rgb}{0.7, 0.1, 0.1}
\usepackage[colorlinks,
    linkcolor=Maroon,
    citecolor=NewBlue,
    urlcolor=NewRed]{hyperref}

\usepackage{cleveref}

\newcommand{\avg}[1]{\left\langle{#1}\right\rangle}

\usepackage{empheq}
\newcommand{\eqdef}{\coloneqq}

\renewcommand{\t}[1]{\mathrm{#1}}

\renewcommand{\phi}{\varphi}

\newcommand{\LigoMIT}{LIGO Laboratory, Massachusetts Institute of Technology, Cambridge, MA 02139}
\newcommand{\MechMIT}{Department of Mechanical Engineering, Massachusetts Institute of Technology, Cambridge, MA 02139}

\begin{document}

\title{Quantum noise and its evasion in feedback oscillators}

\author{Hudson A. Loughlin}
\email{hudsonl@mit.edu}
\affiliation{\LigoMIT}
\author{Vivishek Sudhir}
\affiliation{\LigoMIT}
\affiliation{\MechMIT}

\date{\today}

\begin{abstract}
	We study an abstract model of an oscillator realized by an amplifier embedded in a positive feedback loop. 
	The power and frequency stability of the output of such an oscillator are limited by quantum noise added by two elements in the loop: the amplifier, and the out-coupler. 
	The resulting frequency instability gives the Schawlow-Townes formula. Thus the applicability of
	the Schawlow-Townes formula is extended to a large class of oscillators, and is shown to be related to the 
	Haus-Caves quantum noise limit for a linear amplifier, while identifying the role of quantum noise 
	added at the out-coupler. 
  	By illuminating the precise origin of amplitude and frequency quantum noise in the output of an oscillator, 
  	we reveal several techniques to systematically evade them.
\end{abstract}

\maketitle %

\section{Introduction}

An amplifier embedded in an appropriate positive feedback loop realizes an oscillator \cite{Nyq32,Black34}. 
This basic result of the classical theory of feedback systems \cite{Bode45} has been used as a template for 
designing oscillators that operate in the quantum regime. Famously, the laser \cite{GordTow55,SchaTow58} may 
be considered a regenerative oscillator whose performance can be limited by quantum fluctuations 
\cite{ShimTow57,Pound57,HempLax67,ScuLam67}. In particular, the output frequency noise of a laser or maser is given by the (modified) Schawlow-Townes formula \cite{HempLax67,ScuLam67,Hen83,Pollnau20},
\begin{equation}\label{int:st}
	\bar{S}_{\dot{\phi} \dot{\phi}}^\text{osc}[\Omega] 
	= \frac{\hbar \omega_\text{osc} \kappa^2}{2 P_\text{osc}} \left(1 + 2 n_\text{th} \right) 
	= \frac{\left( \ln(\eta) \right)^2}{2 \tau^2 |\alpha|^2}(1 + 2 n_\text{th}),
\end{equation}
for the symmetrized double-sided spectrum of the frequency deviation $\dot{\phi}$ from the oscillator's nominal output frequency $\omega_\t{osc}$. 
Here, $P_\t{osc} = \hbar \omega_\t{osc}\abs{\alpha}^2$ is the oscillator's mean output power and $\alpha$ is the mean amplitude of the photon flux.
In a laser, the feedback element is a cavity, whose round-trip delay time is $\tau$, average thermal occupation is 
$n_\t{th}=\left[\exp(\hbar\omega_\t{osc}/k_B T)-1\right]^{-1}$, and light is out-coupled through a mirror with power reflectivity $\eta$, equivalent to the cavity linewidth $\kappa = \tau^{-1} \ln \eta$.
The Schawlow-Townes formula is more commonly specified as a linewidth instead of a spectral density. 
The full-width-at-half-maximum linewidth $\Gamma$ of an oscillator with a flat frequency noise spectrum is
given by \cite{Domenico10}, $\Gamma = \bar{S}_{\dot{\phi}\dot{\phi}}/(2\pi)$, so that the linewidth
corresponding to \cref{int:st} is
\begin{equation}\label{int:stLw}
  \Gamma_\text{ST} = \frac{\left( \ln(\eta) \right)^2}{4 \pi \tau^2 |\alpha|^2}(1 + 2 n_\text{th}).
\end{equation}
Note that both \cref{int:st,int:stLw} include the correction for an oscillator above threshold \cite{HempLax67}.
We henceforth confine attention to the performance of quantum-noise-limited oscillators, 
corresponding to the case $n_\t{th}=0$ (we also neglect corrections 
known to arise from coupling of temporal and/or spatial modes \cite{Yariv89}).

We analyze a general model of a feedback oscillator formed by positive feedback of the output of a quantum-noise-limited
amplifier, and show that the Schawlow-Townes formula applies to this general case.
In particular, we identify the precise ingredients of the Schawlow-Townes limit: 
the unavoidable quantum noise added by the amplifier --- the well-known Haus-Caves limit \cite{Hau62,Cav82} --- 
and the noise added by the unavoidable requirement to out-couple the signal from the feedback loop. 
This realization suggests routes to improve the performance of oscillators beyond that 
predicted by the Schawlow-Townes formula:
the conventional, phase-insensitive and thus noisy, amplifier can be replaced with a phase-sensitive noiseless one; 
and/or the quantum noise added at the out-coupler and amplifier can be squeezed or correlated.

Our analysis also derives a general uncertainty principle for the amplitude and phase of the output of a general feedback oscillator.
The Schawlow-Townes formula happens to be one instance of this constraint. 
Another implication is that techniques to evade it based on injection of squeezed vacuum leads to increased fluctuations in the 
amplitude of the output. 
If the feedback oscillator uses a phase-sensitive amplifier instead of a 
phase-insensitive amplifier in its feedback loop or has Einstein-Podolsky-Rosen (EPR) entangled fields, the amplitude and phase 
of the oscillator's output obey less restrictive bounds. 
It is then possible to design oscillators with frequency stability beyond the Schawlow-Townes limit 
without being penalized by increased output power fluctuations.
These results generalize prior work on the Schawlow-Townes formula \cite{HempLax67,ScuLam67,Hen83,Pollnau20,Wise99} and its
evasion \cite{Baker20,Ostrowski23} that relied on a detailed model of laser dynamics.
Our analysis thus clarifies the quantum limits to the stability of a large class of oscillators and
the efficacy of various kinds of quantum resources in improving their performance.

\section{Quantum noise in feedback oscillators}

\begin{figure}[t!]
  \centering
  \includegraphics[width=0.8\columnwidth]{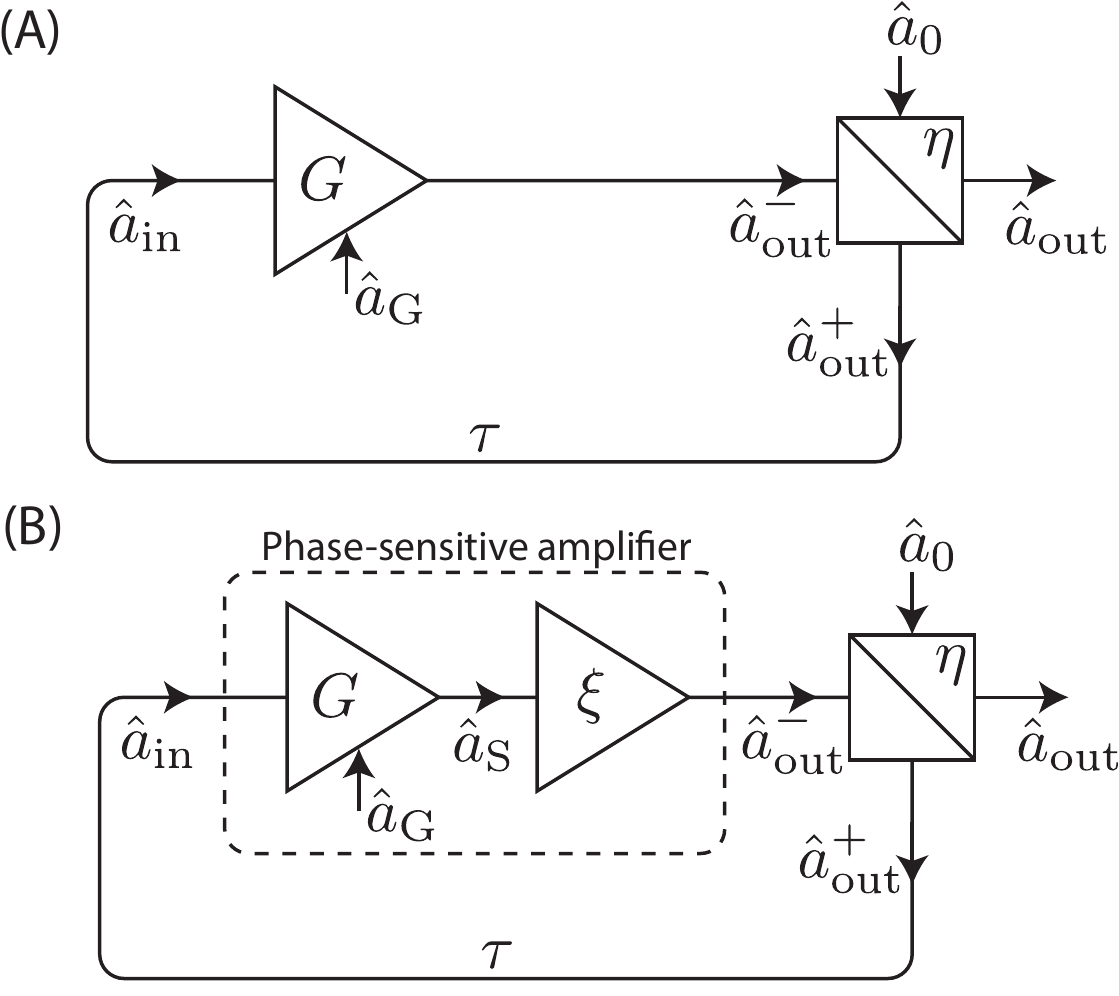}
  \caption{\label{fig:fbamp} (A) A minimal feedback oscillator, which contains a phase-insensitive amplifier embedded in a feedback circuit with a unitary element to couple signal out of the loop. (B) A similar feedback oscillator containing phase-sensitive amplifier (indicated by the dashed rectangle), which is decomposed into a phase-insensitive amplifier followed by a squeezer.}
\end{figure}

We consider (see \cref{fig:fbamp}A) the simplest configuration of a feedback oscillator consistent with the 
laws of quantum physics: %
a phase-insensitive amplifier with \emph{linear} gain $G$ embedded in a feedback loop of time delay $\tau$, 
whose output, because of the no-cloning theorem \cite{WooZur82}, is extracted from the loop using a beam-splitter. 
Clearly, the mode $\hat{a}_0$ adds quantum noise at the out-coupler.
In the absence of the feedback loop, the observation of Haus-Caves \cite{Hau62,Cav82} is that 
the classical input-output relation $\langle\hat{a}^-_\t{out}\rangle = G \avg{\hat{a}_\t{in}}$, cannot be promoted to 
the relation between operators $\hat{a}_\t{out}^-(t) = \hat{a}_\t{in}(t)$, since that is inconsistent
with the commutation relations
(here $i,j \in \{\t{in},\t{out}\}$)
\begin{equation}\label{eq:freeCommute}
  [\hat{a}_i(t),\hat{a}_j^\dagger (t')] = \delta(t-t') \delta_{ij}.
\end{equation} 
This necessitates an ancillary mode $\hat{a}_\t{G}$, with $\avg{a_\t{G}(t)} = 0$, so that the
modified input-output relation,
\begin{equation}\label{eq:HausCaves}
	\hat{a}_\t{out}^-(t) = G \hat{a}_\t{in}(t) + \sqrt{G^2 -1}\, \hat{a}_\t{G}^\dagger (t),	
\end{equation}
is consistent. 
The ancillary mode conveys unavoidable noise added by the amplifier's internal degrees of 
freedom \cite{ScuLam67,Glaub86}.

These observations do not apply when the feedback loop is closed. Firstly, positive feedback will lead to 
a large mean field at the input of the amplifier, which will saturate its output, so that the amplifier cannot be
modeled as being linear. This is a fundamental requirement of any
physical amplifier since its gain arises from an external source, whose energy density has to be finite. 
Secondly, the commutation relations in \cref{eq:freeCommute} only apply to
freely propagating fields, not those inside a feedback loop \cite{Shap87,Wise99b}. 
We will now resolve these issues and show that a proper account of the saturation behavior leads to the
threshold condition for oscillation: ``gain = loss'', while a proper imposition of the commutation 
relation gives the correct quantum noise of the oscillator.

\subsection{Saturation, steady-state, and linear gain}\label{sec:SaturatingFbOsc}

The \emph{nonlinear} input-output behavior of the (memoryless) amplifier can be cast as
\begin{equation}\label{eq:ampOut}
  \alpha_\t{out}^-(t) = \mathcal{A}(\alpha_\t{in}(t)),
\end{equation}
where, $\alpha_i = \avg{\hat{a}_i}$ ($i \in \{\t{in},\t{out}\}$) is the mean amplitude, and $\mathcal{A}(\cdot)$ is the nonlinear gain function
which we postulate has the following properties:
\begin{enumerate}
  \item $\mathcal{A} (-x) = -\mathcal{A} (x)$, i.e. the amplifier is symmetric and bipolar.
  \item $\dv{\mathcal{A}}{x} > 0\quad \forall\, x$, i.e. the gain is monotonic.
  \item $\mathcal{A} (x\rightarrow 0) \rightarrow G_0 x \text{ for some } G_0 > 1/\sqrt{\eta}$, i.e. there exists a ``small signal'' regime 
  of linear gain $G_0$, such that this gain is larger than the loss via the out-coupler, parametrized by $\sqrt{\eta}$.
  \item $\mathcal{A} (x\rightarrow \pm \infty) \rightarrow \pm \mathcal{A}_\infty$, i.e. the output saturates to a constant asymptote for large inputs.
\end{enumerate}

The output of the amplifier $\alpha_\t{out}^- (t)$ propagates through the feedback loop and appears as
\begin{equation}\label{fb:ampLoop}
  \alpha_\t{in}(t) %
    = -\sqrt{\eta}\, \alpha_\t{out}^- (t- \tau) + \sqrt{1-\eta}\, \alpha_0 (t- \tau)
\end{equation}
at the input of the amplifier. 
The classical, nonlinear, input-output relations \cref{eq:ampOut,fb:ampLoop} together with the postulates of 
the nonlinear gain, $\mathcal{A}$, suffice to determine the oscillator's classical steady-state.
As discussed in detail in \Cref{app:saturation}, the oscillator will begin at an unstable equilibrium point with 
zero field amplitude. Noise will then kick the oscillator away from the unstable equilibrium and initiate 
oscillations (just as in a laser \cite{Milonni89}).
Small signal analysis shows that this oscillatory state is stable if
the in-loop field amplitude, $\alpha_\t{ss}$, satisfies
\begin{equation}\label{sat:steadyStateAmp}
	|\mathcal{A}[\alpha_\t{ss}]| = \alpha_\t{ss}/\sqrt{\eta}.
\end{equation}
This output determines the oscillator's amplitude, and serves as the phase reference for quantum fluctuations 
which will be discussed through out this paper. The \emph{linear} gain of the amplifier in this steady-state 
is [\Cref{app:saturation}]
\begin{equation}\label{sat:G}
  G \equiv \lim_{t \to +\infty} \alpha_{\text{out}}^- (t)/\alpha_\text{in} (t) = 1/\sqrt{\eta}.
\end{equation}
This equation has the form ``gain = loss'', and defines the linear gain seen by the
quantum fluctuations on top of the classical steady-state.

\subsection{Quantum fluctuations}

Around the steady state, the (Heisenberg-picture) operators representing the quantum fluctuations satisfy the 
set of linear equations:
\begin{equation}\label{amp:eqsFT}
\begin{split}
	\hat{a}_\text{out}^- [\Omega] &= G \, \hat{a}_\text{in}[\Omega] + \sqrt{G^2 -1} \, \hat{a}_\text{G}^\dagger [\Omega] \\
	\hat{a}_\text{out}^+ [\Omega] &= -\sqrt{\eta}\, \hat{a}_\text{out}^- [\Omega] + \sqrt{1-\eta}\, \hat{a}_0 [\Omega] \\
	\hat{a}_\text{out}[\Omega] &= \sqrt{1-\eta}\, \hat{a}_\text{out}^- [\Omega] +\sqrt{\eta}\, \hat{a}_0[\Omega] \\
	\hat{a}_\text{in} [\Omega] &= e^{i\Omega \tau} \hat{a}_\text{out}^+ [\Omega],
\end{split}
\end{equation}
which we express in terms of their Fourier transforms. 
Here $G$ is the linear gain of the amplifier,
the ancillary mode $\hat{a}_\t{G}$ describes the unavoidable noise associated with amplification,
and the last equation
is the Fourier transform of the time-domain relation, $\hat{a}_\t{in}^-(t)=\hat{a}_\t{out}^+(t-\tau)$, 
expressing the delay in the feedback path.
Note that we take the amplifier's gain to be frequency-independent for frequencies comparable to the 
inverse delay of the loop.

The quantum statistics of the ancillary mode, $\hat{a}_\t{G}$, in closed-loop operation (i.e. $0 < \eta < 1$), 
need not coincide with those in open-loop operation ($\eta = 0$) \cite{Hau62,Cav82,Clerk10}. 
In the closed-loop configuration considered here, the ancillary mode is not freely propagating, 
so it need not satisfy the commutation relations
(the Fourier transform analog of \cref{eq:freeCommute})
\begin{equation}\label{amp:commut}
	[\hat{a}_i [\Omega],\hat{a}_j^\dagger [\Omega']] = 2\pi\cdot \delta [\Omega+\Omega'] \delta_{ij}.
\end{equation}
of a freely propagating field \cite{Shap87,Wise99b}. 
However, the fields in-coupled and out-coupled from the oscillator, $\hat{a}_0$ and $\hat{a}_\t{out}$, are freely propagating, 
and must obey \cref{amp:commut}. 
To enforce this constraint, %
we solve \cref{amp:eqsFT} so that the output can be expressed in terms of the in-coupled 
and ancillary fields:
\begin{equation}\label{amp:sol}
	\hat{a}_\text{out}[\Omega] = H_0[\Omega]\hat{a}_0[\Omega] + H_\text{G} [\Omega] \hat{a}_\text{G}^\dagger [\Omega],
\end{equation}
where 
\begin{equation}\label{amp:H0HG}
\begin{split}
	H_0[\Omega] &= 
		\frac{\sqrt{\eta}+ G e^{i \Omega \tau}}{1 + G \sqrt{\eta} e^{i \Omega \tau}} =
		\frac{\sqrt{\eta}+e^{i \Omega \tau}/\sqrt{\eta}}{1+e^{i \Omega \tau}},\\
	H_\text{G}[\Omega] &= 
		\frac{\sqrt{G^2 - 1} \sqrt{1-\eta}}{1+ G\sqrt{\eta} e^{i \Omega \tau}} =
		\frac{1/\sqrt{\eta} - \sqrt{\eta}}{1+e^{i \Omega \tau}}
\end{split}
\end{equation}
are the transfer functions from the in-coupled and ancillary modes to the output; the second equalities
apply for closed operation in steady-state for which [\cref{sat:G}] $G\sqrt{\eta} = 1$ . 
Note that
\begin{equation}\label{amp:H0HGdiff}
	\abs{H_0[\Omega]}^2 - \abs{H_\text{G}[\Omega]}^2 = 1.
\end{equation}
Insisting that $\hat{a}_\t{out}$ and $\hat{a}_0$ satisfy \cref{amp:commut}, forces
the ancillary mode to satisfy
\begin{equation}\label{ampMode}
\begin{split}
	[\hat{a}_\text{G}[\Omega],\hat{a}_\text{G}^\dagger [\Omega']] =& \frac{|H_0 [\Omega]|^2 - 1}{|H_\text{G} [\Omega]|^2} 2\pi\cdot \delta [\Omega+\Omega']\\
  =& \, 2\pi\cdot \delta [\Omega+\Omega'],
\end{split}
\end{equation}
so although it is not freely propagating, $\hat{a}_\t{G}$ %
obeys the usual canonical commutation relation.

\subsection{Output spectrum and Schawlow-Townes formula}

Our interest is in the amplitude and phase quadratures of the output field. 
For an arbitrary field $\hat{a}_i$, these are defined by $q_i = (\hat{a}_i^\dagger + \hat{a}_i)/\sqrt{2}$ and 
$p_i = i(\hat{a}_i^\dagger - \hat{a}_i)/\sqrt{2}$. 
\Cref{amp:sol} can then be written as
\begin{equation}\label{amp:qpOut}
\begin{split}
  \hat{q}_\text{out}[\Omega] &= H_0[\Omega] \hat{q}_0[\Omega] + H_\text{G} [\Omega] \hat{q}_\text{G}[\Omega] \\ 
  \hat{p}_\text{out}[\Omega] &= H_0[\Omega] \hat{p}_0[\Omega] - H_\text{G} [\Omega] \hat{p}_\text{G}[\Omega]
\end{split}
\end{equation}
Assuming that the in-coupled and ancillary modes are in uncorrelated vacuum states,
i.e. $\bar{S}_{qq}^0 = \bar{S}_{pp}^0 = \bar{S}_{qq}^\text{G} = \bar{S}_{pp}^\text{G} = \frac{1}{2}$ and 
all cross-spectra identically zero (see \Cref{app:sqz}), the output quadrature spectra are
$\bar{S}_{qq}^\text{out} [\Omega] = \bar{S}_{pp}^\text{out} [\Omega] = \frac{1}{2}(\vert H_0^2[\Omega]\vert +\vert H_\text{G}^2[\Omega]\vert)$.
Using the explicit forms of the transfer functions [\cref{amp:H0HG}]
\begin{equation}\label{amp:outQuadSpec}
\begin{split}
  \bar{S}_{qq}^\text{out} [\Omega] = \bar{S}_{pp}^\text{out} [\Omega] 
  = \frac{(\sqrt{\eta}-1/\sqrt{\eta})^2}{4 \cos^2 (\Omega \tau/2)} + \frac{1}{2}.
\end{split}
\end{equation}
As we will now show, the first term is the near-carrier Lorentzian shaped vacuum noise of the oscillator's output ---
whose phase quadrature part is equivalent to the 
Schawlow-Townes formula --- while the second term is the away-from-carrier frequency-independent vacuum noise.

\begin{figure}[t!]
  \centering
  \includegraphics[width=0.9\columnwidth]{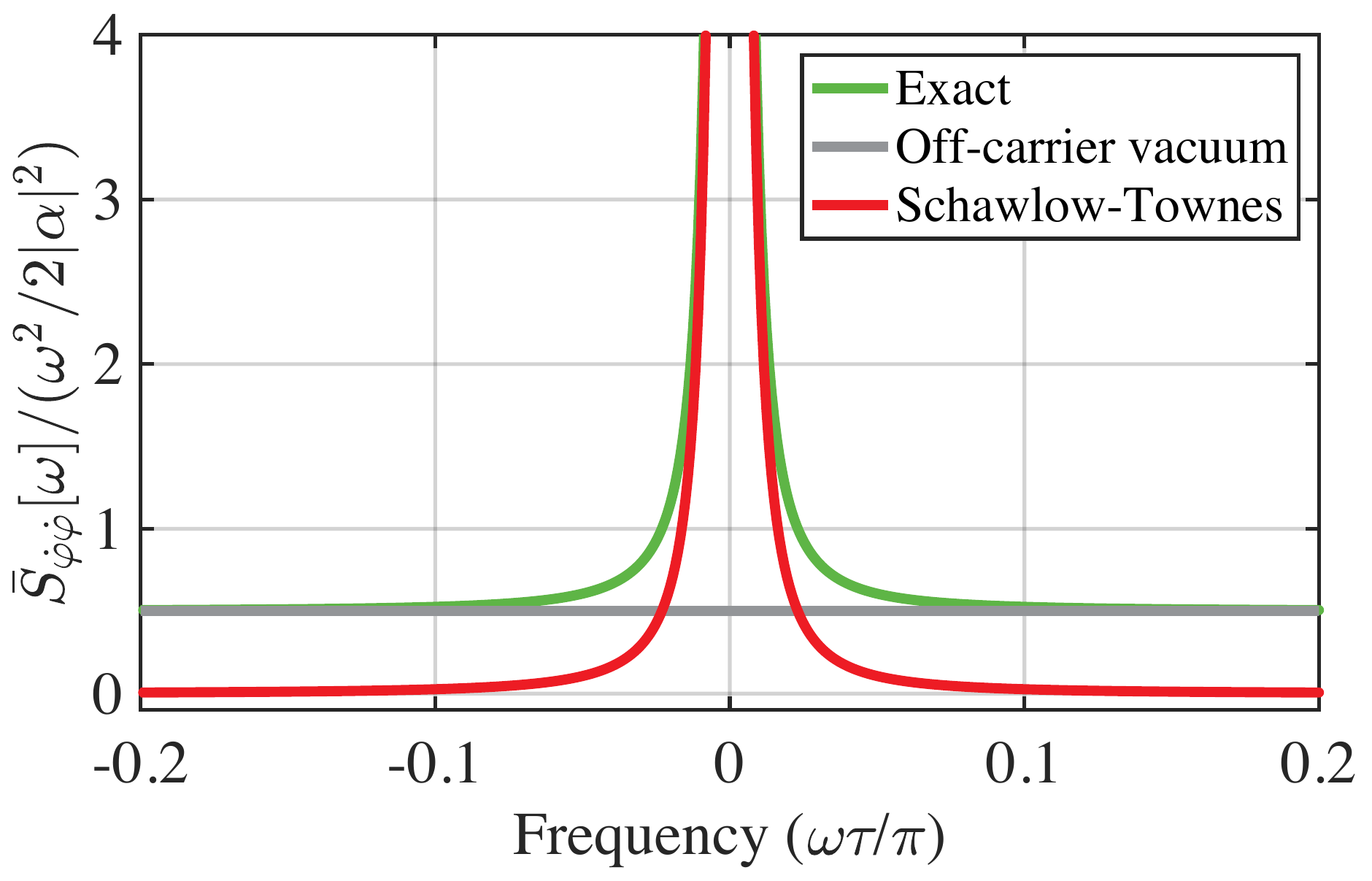}
  \caption{\label{fig:quadSpec} The output phase quadrature spectrum, $\bar{S}_{pp} [\Omega] 
  = \bar{S}_{\dot{\varphi} \dot{\varphi}} [\omega]/(\omega^2/2|\alpha|^2)$, of a feedback oscillator with a 
  phase-insensitive in-loop amplifier is well approximated near the carrier by a second-order pole, producing a 
  Lorentzian line shape. Red shows the Schawlow-Townes component of the lineshape near the carrier, while green
  shows the full prediction [\cref{eq:freqQm}] including the white vacuum noise far from carrier (grey line).}
\end{figure}

The frequencies $\Omega_n \eqdef (2n+1)\pi/\tau$ (for some integer $n$) are the poles of the transfer 
functions $H_{0},H_\t{G}$, 
and are therefore the frequencies of steady-state oscillations of the closed loop. 
Physically, they correspond to constructive interference after subsequent traversals of the feedback loop.
In a practical oscillator, the gain $G[\Omega]$ is typically engineered to sustain only one steady state 
at say frequency $\Omega_0$. 
Then the output field is has a single carrier of amplitude $\alpha = \sqrt{\eta}\alpha_\t{ss}$ that leaks out of the loop.
The above-mentioned quadratures then represent fluctuations around this carrier.
Defining a frequency offset $\omega$ from it, i.e. $\Omega = \Omega_0 + \omega$
where $\omega \tau \ll 1$, the phase quadrature spectrum in \cref{amp:outQuadSpec} assumes the approximate form
\begin{equation}\label{amp:outSpp}
  \bar{S}_{pp}^\t{out}[\Omega_0 + \omega] \approx \frac{(\sqrt{\eta}-1/\sqrt{\eta})^2}{(\omega \tau)^2} +\frac{1}{2}.
\end{equation}
The instantaneous phase deviation of the oscillator's output can be taken to be\footnote{It is well-known 
that the existence of the ground state is a technical obstruction to defining a 
hermitian phase operator, in the full Hilbert space, that is conjugate to the number operator and 
has an eigenvalue spectrum modulo $2 \pi$ \cite{SusGlo64,BarnPeg86}.
For states that have a large mean photon number --- such as the output of an oscillator --- a phase
operator that describes fluctuations about the large carrier field can be defined \cite{HauTow62} 
(see also Appendix G of Ref. \cite{Clerk10}).} $\hat{\phi} \approx \hat{p}_\text{out}/(\sqrt{2} |\alpha|)$.
Thus the frequency spectrum of the output around the carrier is 
$\bar{S}_{\dot{\phi} \dot{\phi}} [\Omega_0 + \omega] 
= \omega^2 \bar{S}_{\phi \phi} [\Omega_0 + \omega] = (\omega^2/2 \abs{\alpha}^2) 
\bar{S}_{pp}^\t{out}[\Omega_0 + \omega]$. 
Using \cref{amp:outSpp}, this is
\begin{equation}\label{eq:freqQm}
\begin{split}
	\bar{S}_{\dot{\phi} \dot{\phi}} [\Omega_0 + \omega] 
	\approx \frac{(\sqrt{\eta} - 1/\sqrt{\eta})^2}{2 \tau^2 \abs{\alpha}^2} 
		+ \frac{(\omega \tau)^2}{4 \tau^2 \abs{\alpha}^2}
	\approx \frac{(1-\eta)^2}{2\tau^2 \abs{\alpha}^2}.
\end{split}
\end{equation}
Here we have omitted the second term, which arises from the frequency-independent vacuum noise ``$\tfrac{1}{2}$''
in \cref{amp:outSpp}, and is negligible close to the carrier. The resulting expression is the Schawlow-Townes 
formula in its regime of applicability
(a laser with a highly reflective out-coupler, i.e. $\eta \approx 1$, in which case \cref{eq:freqQm,int:st} agree 
to $\mathcal{O}[(1-\eta)^3]$). %
\Cref{fig:quadSpec} compares a quantum noise limited oscillator's exact output phase spectrum to the Schawlow-Townes formula and vacuum noise.

\section{Evading the Schawlow-Townes limit}\label{sec:phaseSensFbOsc}

The Schawlow-Townes formula for a quantum-limited feedback oscillator, 
which we henceforth take to be \cref{eq:freqQm}, clearly arises from quantum noises
added by the in-loop amplifier and the out-coupler. We will now consider a fundamental trade-off between
the phase and amplitude fluctuations of the output of a feedback oscillator, illustrate 
how squeezed fields can help evade the Schawlow-Townes limit for the phase 
at the expense of increased power fluctuations, and finally, how entangled fields or a phase-sensitive in-loop amplifier
can evade the fundamental trade-off altogether.

\subsection{An uncertainty relation for feedback oscillators}\label{sec:heis}

The output field quadratures satisfy the canonical commutation relations $[\hat{q}_\t{out}(t),\hat{p}_\t{out}(t')]
= i\, \delta(t-t')$. This fact alone implies that (see \Cref{app:thm:uncertainty})
\begin{equation}\label{unc:SqqSppGen}
	\bar{S}_{qq}^\t{out}[\Omega] \bar{S}_{pp}^\t{out}[\Omega] \geq \frac{1}{4};
\end{equation}
a Heisenberg uncertainty principle for the spectra of these quadratures. This 
constraint is independent of the physics of a feedback oscillator, and is thus too lax.

To derive a tighter constraint, we use \cref{amp:qpOut} 
to relate the output spectra to the in-coupled and ancillary spectra.
For each quadrature $x=\{q,p\}$,
\begin{equation}\label{unc:SqqppOut}
\begin{split}
	\bar{S}_{xx}^\text{out} =& \abs{H_0}^2 \bar{S}_{xx}^0 + \abs{H_\text{G} }^2 \bar{S}_{xx}^\text{G} \pm 2 \text{Re} \left[ H_0^* H_\text{G} \bar{S}_{xx}^{0,\text{G}} \right] \\
		\geq& 2\abs{H_0 H_\text{G} } \left( \sqrt{\bar{S}_{xx}^0 \bar{S}_{xx}^\text{G}} - \abs{\bar{S}_{xx}^{0,\text{G}}} \right), 
\end{split}
\end{equation}
where we have used the notation $\bar{S}_{qq}^\t{0,G} \equiv \bar{S}_{q_0 q_\t{G}}$ etc.
The inequality follows from the fact that $a + b \geq 2\sqrt{ab}$, valid for 
any $a,b \geq 0$, 
and that $-\abs{z} \leq \text{Re}(z) \leq \abs{z}$ for any complex number $z$.
In the first line of \cref{unc:SqqppOut}, the plus sign applies for the amplitude quadrature and the negative sign for the phase quadrature. 
However, the inequality applies for both quadratures.
Using \cref{unc:SqqppOut}, we find that the output quadrature spectra satisfy the constraint
\begin{equation}\label{unc:genUncertainty}
\begin{split}
  \bar{S}_{qq}^\text{out} \bar{S}_{pp}^\text{out} \geq& 4\abs{H_0 H_\text{G} }^2 \times \\
  & \, \left( \sqrt{\bar{S}_{qq}^0 \bar{S}_{qq}^\text{G}} - \abs{\bar{S}_{qq}^{0,\text{G}}} \right) \left( \sqrt{\bar{S}_{pp}^0 \bar{S}_{pp}^\text{G}} - \abs{\bar{S}_{pp}^{0,\text{G}}} \right),
\end{split}
\end{equation}
which applies to all phase-insensitive feedback oscillators.

If further the modes $\hat{a}_0$ and $\hat{a}_\t{G}$ are uncorrelated, 
the cross-terms $\bar{S}_{qq}^{0,\text{G}}$ and $\bar{S}_{pp}^{0,\text{G}}$ vanish.
Because the modes $\hat{a}_0$ and $\hat{a}_\t{G}$ satisfy the canonical commutation relations, 
it follows that their quadratures satisfy the uncertainty principle (see \Cref{app:thm:uncertainty})
$\bar{S}_{qq}^i \bar{S}_{pp}^i \geq \tfrac{1}{4}$ for each mode $i=\{0,G\}$.
Thus \cref{unc:genUncertainty} reduces to
\begin{equation}\label{unc:noCorr}
	\bar{S}_{qq}^\text{out} \bar{S}_{pp}^\text{out} \geq \abs{H_0 H_\text{G}}^2 
	= \abs{H_0}^2 \left(\abs{H_0}^2 - 1 \right).
\end{equation}
This is a \emph{state-independent} constraint on the trade-off in the 
fluctuations in the output field of a feedback oscillator formed by positive feedback of a phase-insensitive amplifier 
in the absence of quantum correlations between its in-coupled and ancillary modes. 

Around the carrier, where $\abs{H_0} \gg 1$, \cref{unc:noCorr} is much tighter than the Heisenberg uncertainty
principle in \cref{unc:SqqSppGen}. Indeed, the Schawlow-Townes limit is an instance of \cref{unc:noCorr}.
To wit, note that for
frequencies near the carrier, $\abs{H_0}~\gg~1$, so \cref{unc:noCorr} implies 
$\bar{S}_{qq}^\t{out} \bar{S}_{pp}^\t{out} \gtrsim \abs{H_0}^4$. 
The Schawlow-Townes limit is the case where this inequality is saturated by an equal partitioning
of fluctuations between the two output quadratures, i.e. $\bar{S}_{qq}^\t{out} \approx \bar{S}_{pp}^\t{out} 
\approx \abs{H_0}^2$.
This means that as long as the modes $\hat{a}_0,\hat{a}_\t{G}$ are independent, and the amplifier is
phase-insensitive, any attempt to reduce frequency fluctuations below the Schawlow-Townes 
limit --- by engineering the out-coupler or ancillary states --- will elicit increased 
fluctuations in the output power of the oscillator.
Any and all of these assumptions can be relaxed to evade the Schawlow-Townes limit to varying degrees of
malleability.
In fact, it is apparent that correlating the modes $\hat{a}_0,\hat{a}_\t{G}$ weakens the uncertainty bound in \cref{unc:genUncertainty}.

\subsection{Phase-insensitive in-loop amplifier: squeezing and entanglement}\label{sec:sqzEpr}

We now consider in concrete terms the possibility of evading the Schawlow-Townes limit for 
a feedback oscillator made of a phase-insensitive in-loop amplifier. 
In order to focus on the quadrature fluctuations near the carrier, we express \cref{amp:qpOut}
at the offset frequency $\Omega_0 + \omega$
\begin{equation}\label{heis:qpOut}
\begin{split}
	\hat{q}_\text{out}[\omega] &\approx H_0[\omega] \left( \hat{q}_0[\omega] - \hat{q}_\text{G}[\omega] \right)\\ 
	\hat{p}_\text{out}[\omega] &\approx H_0[\omega] \left( \hat{p}_0[\omega] + \hat{p}_\text{G}[\omega] \right).
\end{split}
\end{equation}
For brevity, here and henceforth, we write $\omega$ in lieu of $\Omega_0 + \omega$.
Then the spectrum of the output phase and amplitude quadratures assumes the general form 
\begin{equation}\label{SppoutSpectralRelation}
\begin{split}
	\Bar{S}_{qq}^\text{out} [\omega] =& \abs{H_0 [\omega]}^2 \left( \Bar{S}_{qq}^0 [\omega]
		+ \Bar{S}_{qq}^\text{G} [\omega] - 2 \, \text{Re} \left[\bar{S}_{qq}^{0,\text{G}} [\omega] \right] \right)\\
	\Bar{S}_{pp}^\text{out}[\omega] =& \abs{H_0 [\omega]}^2 \left( \Bar{S}_{pp}^0 [\omega]
		+ \Bar{S}_{pp}^\text{G}[\omega] + 2 \, \text{Re} \left[ \bar{S}_{pp}^{0,\text{G}}[\omega] \right] \right).
\end{split}
\end{equation}

Clearly, squeezing either the in-coupled or ancillary mode can reduce the noise power
in a desired quadrature. Entangling these modes --- resulting in non-zero
$\bar{S}_{qq}^\t{0,G}$ and $\bar{S}_{pp}^\t{0,G}$ --- can reduce fluctuations in both quadratures simultaneously.
For illustration, we imagine frequency-independent single-mode squeezing of the in-coupled mode $\hat{a}_0$ and the 
amplifier's ancillary mode $\hat{a}_\t{G}$, 
with squeezing parameters $r_0$ and $r_\t{G}$ respectively, 
followed by two-mode squeezing of the modes $\hat{a}_0$ and $\hat{a}_\t{G}$
with squeezing parameter $r_\text{E}$. 
(The latter corresponds to continuous-variable EPR entanglement of the two modes 
$\hat{a}_0$ and $\hat{a}_\t{G}$ \cite{BrauKimb98}.)
Gaussian state techniques allow the resulting output spectra to be derived (see \Cref{app:sqz} for details):
\begin{equation}\label{eq:sqzCorrSpec}
\begin{split}
  \bar{S}_{qq}^\text{out} [\omega] &= \frac{1}{2} e^{-r_\text{E}} \left( e^{2 r_0} + e^{2 r_\text{G}} \right) \abs{H_0[\omega]}^2 \\
  \bar{S}_{pp}^\text{out} [\omega] &= \frac{1}{2} e^{-r_\text{E}} \left( e^{-2 r_0} + e^{-2 r_\text{G}} \right) \abs{H_0[\omega]}^2
\end{split}
\end{equation}
By squeezing their phase-quadratures, corresponding to $r_0, r_\t{G} > 0$,
we can suppress the oscillator's phase quadrature fluctuations --- and therefore the linewidth of the oscillator --- 
without bound, at the expense of increasing the oscillator's amplitude quadrature fluctuations.
By correlating these modes, corresponding to $r_\t{E} > 0$, we can simultaneously reduce the oscillator's
amplitude and phase quadrature fluctuations. However, the Heisenberg uncertainty relation [\cref{unc:SqqSppGen}]
$\bar{S}_{qq}^\t{out} \bar{S}_{pp}^\t{out}\geq \tfrac{1}{4}$ still holds, and 
represents the limit to which noise in the output quadratures of a feedback oscillator can be suppressed
(see \Cref{app:sqz} for an explicit verification of this fact for EPR entangled inputs).

\subsection{Phase-sensitive in-loop amplifier}\label{sec:phaseSensAmp}

An alternative method of reducing the oscillator's output amplitude and phase spectra is to modify the feedback loop itself by 
replacing the phase-insensitive amplifier by a phase-sensitive amplifier as depicted in \Cref{fig:fbamp}(B). 
For the sake of generality, we consider that the phase-sensitive amplifier is realized by phase-insensitive amplification of the
output of an ideal (i.e. noiseless) phase-sensitive amplifier. The latter is a squeezer with squeezing parameter $\xi = r_\t{s}e^{i\phi_\t{s}}$.
When the phase-insensitive component has unity-gain, this cascade realizes a noiseless phase-sensitive amplifier (see \Cref{sec:phaseSensitiveAmpDecomp}).
As before, the nonlinear saturating response of the phase-insensitive amplifier limits the oscillator's output; a straightforward extension of the
prior analysis shows that this happens when $G e^{r_\t{s}} \sqrt{\eta} = 1$, which can be interpreted as a balance between gain and loss in the loop, except
that now the gain ($G e^{r_\t{s}}$) is phase-sensitive.
Linear response around this steady-state oscillation is described by
\begin{equation}\label{amp:eqsFtPhaseSensitive}
\begin{split}
	\hat{a}_\text{out}^- [\Omega] &= \cosh (r_\text{s}) \, \hat{a}_\text{s} [\Omega] + e^{i \phi_\text{s}} \sinh (r_\text{s}) \, \hat{a}_\text{s}^\dagger [\Omega] \\
	\hat{a}_\text{s} [\Omega] &= G \, \hat{a}_\text{in}[\Omega] + \sqrt{G^2 - 1} \, \hat{a}_\text{G}^\dagger[\Omega] \\
	\hat{a}_\text{out}^+ [\Omega] &= -\sqrt{\eta}\, \hat{a}_\text{out}^- [\Omega] + \sqrt{1-\eta}\, \hat{a}_0 [\Omega] \\
	\hat{a}_\text{out}[\Omega] &= \sqrt{1-\eta}\, \hat{a}_\text{out}^- [\Omega] + \sqrt{\eta}\, \hat{a}_0[\Omega] \\
	\hat{a}_\text{in} [\Omega] &= e^{i\Omega \tau} \hat{a}_\text{out}^+ [\Omega].
\end{split}
\end{equation}
where, as before, $\hat{a}_0,\hat{a}_\t{G}$ are the modes conveying noise at the out-coupler and the in-loop amplifier.
These equations can be solved for the output quadrature fluctuations in terms of the two noise inputs:
\begin{equation}
\begin{split}
	\hat{q}_\text{out} [\Omega] &= H_0^q [\Omega] \hat{q}_0 [\Omega] + H_\text{G}^q [\Omega] \hat{q}_\text{G} [\Omega] \\
	\hat{p}_\text{out} [\Omega] &= H_0^p [\Omega] \hat{p}_0 [\Omega] - H_\text{G}^p [\Omega] \hat{p}_\text{G} [\Omega].
\end{split}
\end{equation}
Note that the output quadratures are defined with respect to the squeezing angle $\phi_\t{s}$. The full expressions for the four
transfer functions $H_{0,G}^{q,p}$ are given in \Cref{app:phSensFbOsc}. 
Importantly, the noise properties of the amplifier are determined by the response of the feedback loop, which we analyze, as before, 
by demanding that the in-coupled and out-coupled fields satisfy the canonical commutation relations.
This gives (see \Cref{app:phSensFbOsc})
\begin{equation}\label{eq:aGphaseInsensCommutator}
\begin{split}
	[\hat{a}_\text{G} [\Omega], \hat{a}_\text{G}^\dagger [\Omega^\prime]] = 2\pi \cdot \delta [\Omega + \Omega^\prime],
\end{split}
\end{equation}
and we again find that the ancillary mode obeys the canonical commutation relations. 

\begin{figure}[t!]
  \centering
  \includegraphics[width=0.9\columnwidth]{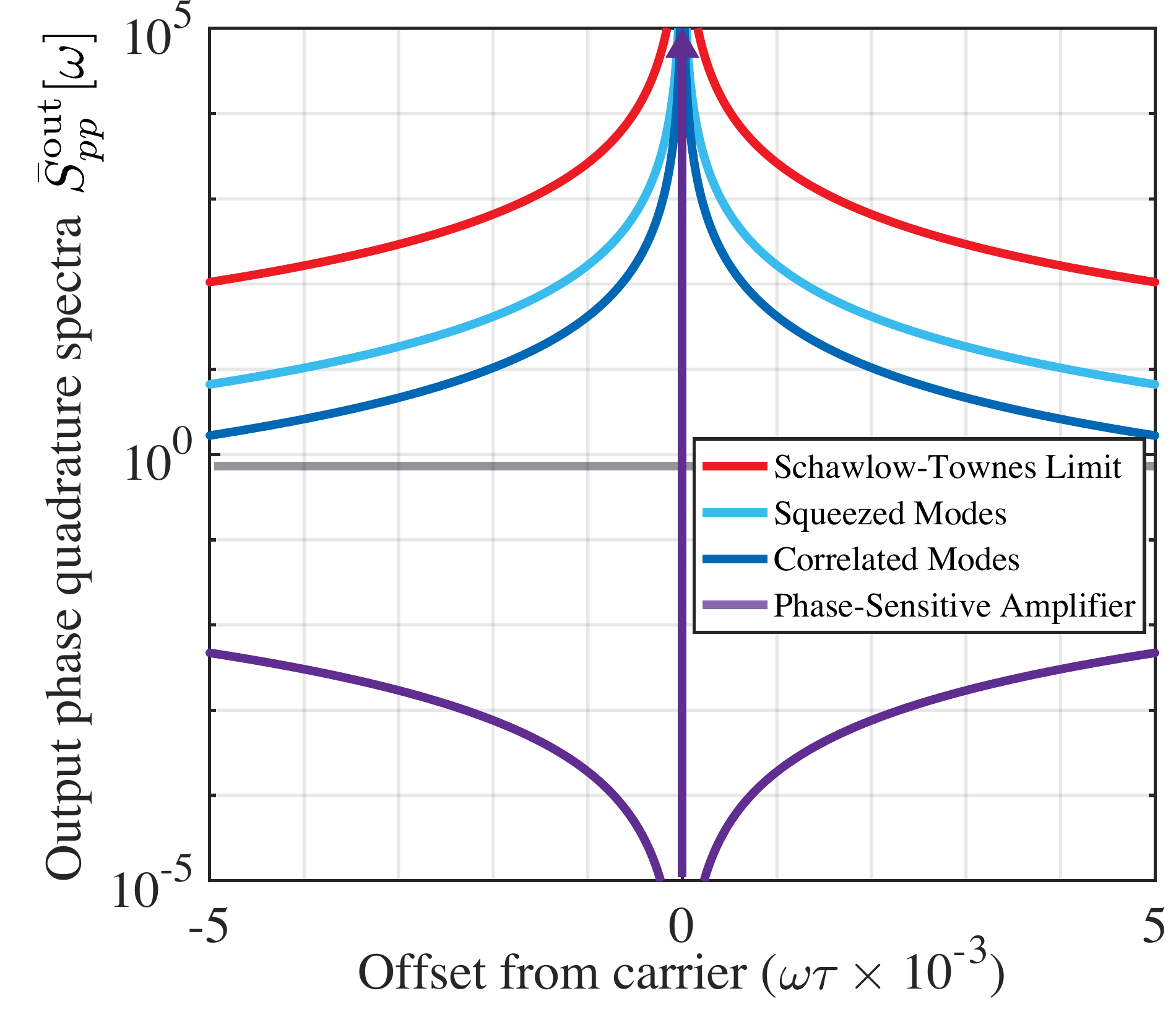}
  \caption{\label{fig:spectra} Spectra of the output phase quadrature for four types of quantum noise limited 
  oscillators. Red shows the Schawlow-Townes spectrum of an oscillator with phase-insensitive
  amplifier and the in-coupled and ancillary modes in vacuum . Light and darker blues depict the case where these
  modes are squeezed (light blue) and entangled (dark blue) (both with 12 dB of squeezing).
  Purple shows the case where the in-loop amplifier is purely phase-sensitive.
  Grey shows $\bar{S}_{pp}^\t{out}=1/2$ for reference.}
\end{figure}

For a feedback oscillator with a phase-sensitive amplifier and vacuum state in-coupled and ancillary modes, the output amplitude and phase quadrature spectra are given by
\begin{equation}\label{eq:SppPhSens}
\begin{split}
	\bar{S}_{qq}^\text{out} [\Omega] &= \frac{1}{2} |H_0 [\Omega]|^2 + \frac{1}{2} \left(\frac{1 - \eta \, e^{2 r_\text{s}}}{1- \eta} \right) |H_\text{G} [\Omega]|^2, \\
	\bar{S}_{pp}^\text{out} [\Omega] &= \frac{1}{2}|H_0^p [\Omega]|^2 + \frac{1}{2} |H_\text{G}^p [\Omega]|^2 .
\end{split}
\end{equation}
In the first equation, we have made use of the fact that $H_0^q = H_0$ and $H_\t{G}^q = \sqrt{(1 - \eta \, e^{2 r_\text{s}})/(1 - \eta)} H_\t{G}$.
We note that $|H_0^p [\Omega]| < |H_0 [\Omega]|$ and $|H_\text{G}^p [\Omega]| < |H_\text{G} [\Omega]|$ near resonance for $0 < r_\t{s} \leq - \ln (\eta)/2$, so using a phase-sensitive amplifier reduces the oscillator's phase quadrature spectrum. (Gain-loss balance requires $r_\t{s} \leq - \ln (\eta)/2$, since the gain of a purely phase-sensitive amplifier is $e^{r_\t{s}}$ and the loss from the out-coupler is $\sqrt{\eta}$). Additionally, the phase-sensitive oscillator's output amplitude quadrature spectrum is always less than that of a phase-insensitive oscillator, although by at most a factor of $1/2$.

Clearly, an oscillator based on feedback of a phase-sensitive amplifier provides a means of evading the Schawlow-Townes limit. The extreme version of this case is where the in-loop amplifier is perfectly phase-sensitive ($G=1$ 
in \cref{amp:eqsFtPhaseSensitive}), i.e. it is a single mode squeezer with no phase-insensitive gain, and thus
noiseless in-loop amplification.
In this case, taking the mode $\hat{a}_0$ to be vacuum, 
the output quadrature spectra around the carrier are (see \Cref{app:phSensFbOsc})
\begin{equation}\label{eq:purePhSens}
\begin{split}
  \bar{S}_{qq}^\text{out} [\omega] = \frac{(\sqrt{\eta}-1/\sqrt{\eta})^2}{\omega^2 \tau^2} \bar{S}_{qq}^0 [\omega] \\
  \bar{S}_{pp}^\text{out} [\omega] = \frac{\omega^2 \tau^2}{(\sqrt{\eta}-1/\sqrt{\eta})^2} \bar{S}_{pp}^0 [\omega].
\end{split}
\end{equation}
Since $\bar{S}_{qq}^\text{out} \bar{S}_{pp}^\text{out} = \bar{S}_{qq}^0  \bar{S}_{pp}^0$, 
the feedback oscillator does not increase the uncertainty product between the amplitude and phase quadratures from its input to output. In particular, if the in-coupled field $\hat{a}_0$ is vacuum, the oscillator's output field is 
a minimal uncertainty squeezed state around the steady-state oscillating carrier --- a bright squeezed state.
(In practice, any technical phase noise in the phase-sensitive amplifier will inject noise from the anti-squeezed 
amplitude quadrature into the squeezed phase quadrature \cite{Dwyer13}, precluding the possibility of 
zero phase noise predicted by \cref{eq:purePhSens} as $\omega \rightarrow 0$).

\section{Discussion and Conclusion}

\Cref{fig:spectra} compares the output phase quadrature spectra of oscillators with squeezed in-coupled and ancillary modes, squeezed and correlated in-coupled and ancillary modes, and an oscillator with a phase-sensitive amplifier. As can be seen in the figure, embedding a phase-sensitive amplifier in the feedback loop suppresses the oscillator's phase-quadrature spectrum much more than is possible by squeezing or correlating the input modes for a given maximal level of squeezing. 
We also emphasize that unlike squeezing the in-coupled and ancillary modes, correlating these modes or embedding a phase-sensitive amplifier in the feedback loop reduces the oscillator's phase quadrature spectrum without increasing its amplitude quadrature spectrum.

We have identified the origin of the Schawlow-Townes limit to the frequency stability of an oscillator in a manner
that is applicable to a wide class of feedback oscillators. In fact, for a phase-insensitive, quantum-noise-limited oscillator, 
it is one facet of a more general uncertainty principle for the oscillator's outgoing field. The constraint posed by
this uncertainty principle dictates the fundamental trade-off between the frequency and power fluctuations of a broad class of
oscillators. However, systematic strategies such as injection of squeezed vacuum, EPR entanglement, and phase-sensitive amplification offer sufficient
room to evade the Schawlow-Townes limit.

\bibliographystyle{apsrev4-2}
\bibliography{ref_feedback_oscillator.bib}

\appendix

\section{Spectral densities: definition and properties}\label{app:spectra}

For a time-dependent operator (not necessarily hermitian) $\hat{A}(t)$, we define its Fourier transform as
\begin{equation}\label{def:FT}
	\hat{A}[\Omega] = \int_{\mathbb{R}} \hat{A}(t) e^{i\Omega t}\, \dd t,
\end{equation}
where $\Omega \in \mathbb{R}$. Note that the Fourier transform of the hermitian conjugate, which
we denote by $\hat{A}^\dagger [\Omega]$, is different from the hermitian conjugate of the Fourier transform,
which we represent by $\hat{A}[\Omega]^\dagger$; the two are related as,
\begin{equation}\label{eq:FTsymm}
	\hat{A}^\dagger [\Omega] = \hat{A}[-\Omega]^\dagger.
\end{equation}
The inverse of \cref{def:FT} is given by,
\begin{equation}\label{def:IFT}
	\hat{A}(t) = \int_\mathbb{R} \hat{A}[\Omega] e^{-i\Omega t} \frac{\dd \Omega}{2\pi}.
\end{equation}

We define the cross-correlation between the two general (not necessarily hermitian and not necessarily commuting)
operators $\hat{A},\hat{B}$ by the symmetrized expression,
\begin{equation*}
	\bar{S}_{AB}(t) = %
	\frac{1}{2}\avg{\{\hat{A}^\dagger (t),\hat{B}(0)\}}.
\end{equation*}
We will primarily use the \emph{symmetrized double-sided} cross-correlation spectrum, defined as
the Fourier transform of the symmetrized cross-correlation \cite{Clerk10}, viz.,
\begin{equation}\label{app:SABsym1}
\begin{split}
	\bar{S}_{AB}[\Omega] &= \int_{-\infty}^{\infty} \frac{1}{2}\avg{\{\hat{A}^\dagger(t),\hat{B}(0)\}} e^{i\Omega t}\, \dd t \\
		&= \int_{-\infty}^{\infty} \frac{1}{2}\avg{\{\hat{A}^\dagger [\Omega], \hat{B}[\Omega']\}} \frac{\dd \Omega'}{2\pi},
\end{split}
\end{equation}
where the last equality follows from using the Fourier representation, \cref{def:IFT}, of the time-dependent operators.

For weak-stationary operators, i.e. those that pair-wise satisfy
\begin{equation*}
	\avg{\hat{A}(t)\hat{B}(t')} = \avg{\hat{A}(t-t') \hat{B}(0)},
\end{equation*}
the spectrum is given by the identity,
\begin{equation}\label{app:SABsym2}
	\bar{S}_{AB}[\Omega]\cdot 2\pi \delta[\Omega+\Omega'] = \frac{1}{2}\avg{\{\hat{A}^\dagger[\Omega], \hat{B}[\Omega']\}}.
\end{equation}
Note that in this case, $\bar{S}_{AB}[\Omega]^* = \bar{S}_{BA}[\Omega]$.
\begin{lemma}\label{app:lem:posSpec}
	The spectrum of a weak-stationary (but not necessarily Hermitian) operator is positive; i.e.,
	\begin{equation}\label{eq:AdagAGeqZero}
		\bar{S}_{AA}[\Omega] \geq 0.
	\end{equation}
	if $\langle\hat{A}^\dagger (t) \hat{A}(t')\rangle =\langle \hat{A}^\dagger(t-t')\hat{A}(0)\rangle$.
	\begin{proof}
		Since $\hat{A}$ is weak-stationary, \cref{app:SABsym2}, together with \cref{eq:FTsymm}, 
		implies that the spectrum $\bar{S}_{AA}[\Omega]$ satisfies,
		\begin{equation*}
			\bar{S}_{AA}[\Omega]\cdot 2\pi \delta[0] = 	\frac{1}{2}\avg{\{\hat{A}[-\Omega]^\dagger, \hat{A}[-\Omega]\}}.
		\end{equation*}
		Next we prove the right-hand side is positive. Consider the first term, which is the expectation of the hermitian
		operator, $\hat{A}[-\Omega]^\dagger \hat{A}[-\Omega]$ over some state, say $\hat{\rho}$. Since the general state $\hat{\rho}$ can be expressed
		as a convex combination, $\hat{\rho} = \sum_i p_i \ket{\psi_i}\bra{\psi_i}$, with $p_i\geq 0$, $\sum_i p_i =1$, and
		$\langle \psi_i \vert \psi_j\rangle = \delta_{ij}$, the
		expectation value may be written as,
		\begin{equation*}
		\begin{split}
			\avg{\hat{A}[-\Omega]^\dagger \hat{A}[-\Omega]} &= \text{Tr}[\hat{A}[-\Omega]^\dagger \hat{A}[-\Omega] \hat{\rho}] \\
				&= \sum_i p_i \bra{\psi_i}\hat{A}[-\Omega]^\dagger \hat{A}[-\Omega]\ket{\psi_i} \\
				&= \sum_i p_i \| \hat{A}[-\Omega]\ket{\psi_i}\|^2 \geq 0.
		\end{split}
		\end{equation*}
		The same argument applies to the second term.
	\end{proof}
\end{lemma}
We may thus interpret the spectrum of a weak-stationary operator, $\bar{S}_{AA}[\Omega]$, as the variance of the
operator-valued process $\hat{A}(t)$ per unit bandwidth, similar to a classical spectral density.

\begin{theorem}\label{app:thm:uncertainty}
	For for weak-stationary observables $\hat{A}$ and $\hat{B}$ that satisfy the commutation relationship $[\hat{A}[\Omega], \hat{B}[\Omega^\prime]] = i c \cdot 2 \pi \delta[\Omega + \Omega^\prime]$ for some real constant $c \in \mathbb{R}$, 
	\begin{equation}\label{eq:robSchIneq}
		\bar{S}_{A A} [\Omega] \bar{S}_{B B} [\Omega] \geq  \frac{c^2}{4} + \left| \bar{S}_{A B}[\Omega] \right|^2.
	\end{equation}
\end{theorem}

\begin{proof}
	Let $\{ \hat{A}_j \}$ be a set of weak-stationary observables; then their linear combination, 
	$\hat{M} = \sum_j \alpha_j \hat{A}_j$ with $\alpha_j \in \mathbb{C}$ is also weak-stationary. 
	Using the fact that $\langle \hat{M}[-\Omega]^\dagger \hat{M}[-\Omega] \rangle \geq 0$ 
	as shown in the proof of \cref{eq:AdagAGeqZero}, we have
	\begin{equation}
		\text{Tr} \left[ \sum_{j,k} \alpha_j^* \hat{A}_j [\Omega] \alpha_k \hat{A}_k [-\Omega] 
		\hat{\rho} \right] \geq 0,
	\end{equation}
  where we have used \cref{eq:FTsymm} and the fact that $\hat{A}_k^\dagger [\Omega] = \hat{A}_k [\Omega]$ since $\hat{A}_k$ is an observable by assumption.
	Equivalently,
	\begin{equation}\label{eq:covInequality}
		\sum_{j,k} \alpha_j^* \alpha_k \langle \hat{A}_j [\Omega] \hat{A}_k [-\Omega] \rangle \geq 0.
	\end{equation}
	We can split $\hat{A}_j [\Omega] \hat{A}_k [-\Omega]$ into Hermitian and anti-Hermitian parts as
	\begin{equation}\label{eq:splitCommAntiComm}
	\begin{split}
		\hat{A}_j [\Omega] \hat{A}_k [-\Omega] =& \frac{1}{2} \left\{ \hat{A}_j [\Omega] 
		,\hat{A}_k [-\Omega] \right \}  \\
		&+ \frac{1}{2} \left[ \hat{A}_j [\Omega], \hat{A}_k [-\Omega] \right].
	\end{split}
	\end{equation}
	If we define 
	\begin{equation}\label{eq:commutatorSpectrum}
		\bar{C}_{A B} [\Omega] \equiv \int_{-\infty}^{\infty} \frac{1}{2} \left[ \hat{A}^\dagger [\Omega], 
		\hat{B}[\Omega]^\prime \right] \frac{\dd \Omega^\prime}{2 \pi},
	\end{equation}
	then, using \cref{eq:splitCommAntiComm,eq:covInequality,app:SABsym2,eq:commutatorSpectrum} we have
	\begin{equation}\label{eq:commAntiCommMatrixIneq}
		\sum_{j,k} \alpha_j^* \alpha_k \left( \bar{S}_{A_j,A_k}[\Omega] + \bar{C}_{A_j,A_k} [\Omega] \right) \geq{0}.
	\end{equation}
	This inequality requires that the eigenvalues of the matrix with elements given by 
	$\bar{S}_{A_j,A_k}[\Omega] + \bar{C}_{A_j,A_k}[\Omega]$ be real and non-negative. 
	Specifically, the smallest eigenvalue of this matrix must be non-negative.
	
	Consider the case where $\{ \hat{A}_j \} = \{ \hat{A}, \hat{B} \}$ with $[\hat{A}[\Omega], \hat{B}[\Omega^\prime]] 
	= i c \cdot 2 \pi \delta[\Omega + \Omega^\prime]$; here $c \in \mathbb{R}$ by virtue of $\hat{A},\hat{B}$ being observables.
	In this case \cref{eq:commAntiCommMatrixIneq} requires that (dropping frequency arguments for brevity)
	\begin{equation*}
		\bar{S}_{A A} \bar{S}_{B B} - \left(\bar{S}_{A B} + \bar{C}_{A B} \right) 
		\left( \bar{S}_{B A} + \bar{C}_{B A} \right) \geq 0.
	\end{equation*}
	Simplifying using the fact that $\bar{C}_{A B} [\Omega] = \bar{C}_{B A}^* [\Omega] = i c / 2$ and
	$\bar{S}_{AB}[\Omega] = \bar{S}_{BA}^*[\Omega]$, gives \cref{eq:robSchIneq}.
\end{proof}

\section{Details of feedback oscillator based on phase-insensitive amplifier}\label{app:PhaseInsensitive}

\subsection{Saturating behavior and classical steady state}\label{app:saturation}

Here, we analyze the classical steady-state behavior of the saturating feedback oscillator shown 
in \cref{fig:fbamp}A. 
Classically, the behavior of the system shown in \cref{fig:fbamp}A is governed in the time domain by
\begin{equation}\label{amp:eqsClassTime}
\begin{split}
  \alpha_\text{out}^- (t) &= \mathcal{A} [\alpha_\text{in}(t)] \\
  \alpha_\text{out}^+ (t) &= -\sqrt{\eta}\, \alpha_\text{out}^- (t) + \sqrt{1-\eta}\, \alpha_0 (t) \\
  \alpha_\text{out}(t) &= \sqrt{1-\eta}\, \alpha_\text{out}^- (t) +\sqrt{\eta}\, \alpha_0(t) \\
  \alpha_\text{in} (t) &= \alpha_\text{out}^+ (t - \tau).
\end{split}
\end{equation}
where $\{\alpha_i \}$ are classical field amplitudes and $\mathcal{A} [\cdot]$ is the amplifier's nonlinear response. 
\Cref{amp:eqsClassTime} can be simplified to 
\begin{equation}\label{amp:classSat}
  \alpha_\text{out}^+ (t) = -\sqrt{\eta}\, \mathcal{A} [\alpha_\text{out}^+(t-\tau)] + \sqrt{1-\eta}\, \alpha_0 (t-\tau)
\end{equation}
Clearly, the input $\alpha_0(t)$ sources the classical field that circulates in the loop.
In reality, the input represented by $\alpha_0$ is pure noise, in the ideal case, just vacuum noise. 

To analyze how the feedback loop attains a steady state by being driven purely by noise, we assume that
$\alpha_0$ represents infinitesimally small fluctuations. In order to understand how the loop starts, 
consider that $\alpha_0(0)$ is a small random value and zero for $t > 0$. Let this produce a small amplitude
$\alpha_\t{out}^+ (0<t<\tau) = \delta$. 
We are primarily interested in the circulating power rather than phase rotations, so taking the magnitude square
of \cref{amp:classSat} under these conditions:
\begin{equation}\label{amp:classSatFinal}
\begin{split}
  |\alpha_\text{out}^+ (t)|^2 &= \eta\, \left|\mathcal{A} [\alpha_\text{out}^+(t - \tau)]\right|^2 \\
  \alpha_\text{out}^+ (0<t<\tau) &= \delta.
\end{split}
\end{equation}
That is, if the initial random seed $\alpha_0(0)$ produces a steady state amplitude, 
it must satisfy 
\begin{equation}\label{amp:classSatSS}
	\abs{\mathcal{A}[\alpha_\t{out}^+]} = \alpha_\t{out}^+/\sqrt{\eta},
\end{equation}
which defines the steady state $\alpha_{ss}$ of \cref{sat:steadyStateAmp}.

The above is only a necessary condition, since the question of the stability of this steady state remains open.
We will now show that the properties of $\mathcal{A}$ listed in \Cref{sec:SaturatingFbOsc} suffice to ensure
stability.
Properties (1), (3), and (4) imply that $\mathcal{A}^\prime(x) = \frac{d \mathcal{A}}{dx} < \frac{\mathcal{A} (x)}{x} 
\quad \forall x$. Let $\alpha_\t{out}^+(t) = \alpha_{ss} + \delta_t$ where $\delta_t$ is a small 
perturbation to the steady state value of $\alpha_\t{out}^+$ at time $t$. 
Similarly, let $\alpha_\t{out}^+(t + \tau) = \alpha_{ss} 
+ \delta_\tau$. \Cref{amp:classSatFinal} then reads $|\alpha_{ss} + \delta_\tau|^2 = 
\eta|\mathcal{A}[\alpha_{ss}+\delta_t]|^2$. Expanding to first order in $\delta_t,\delta_\tau$ and using
\cref{amp:classSatSS} gives
\begin{equation}
\begin{split}
  |\delta_\tau| &= \frac{\alpha_{ss} \mathcal{A}^\prime[\alpha_{ss}]}{\mathcal{A}[\alpha_{ss}]} |\delta_t| \\
  \Rightarrow\qquad |\delta_\tau| &< |\delta_t|
\end{split}
\end{equation}
where we have used the properties of $\mathcal{A}$. Thus, perturbations around the steady state tend to die over time;
i.e. the steady state is stable to small perturbations.

Similarly, we can show that the operating point where all amplitudes are zero is unstable. That is, any initial 
fluctuation will drive the loop to its steady state \cref{amp:classSatSS}.
Now, let $\alpha_\t{out}^+(t) = \delta_t$ and let $\alpha_\t{out}^+(t + \tau) = \delta_\tau$ where $\delta_t$ and $\delta_\tau$ are sufficiently small that property (2) is satisfied. We have
\begin{equation}
\begin{split}
  |\delta_\tau|^2 &= \eta|\mathcal{A}[\delta_t]|^2 \\
  \Rightarrow \qquad |\delta_\tau| &= \sqrt{\eta} |G_0| |\delta_t| \\
  \Rightarrow \qquad |\delta_\tau| &> |\delta_t|.
\end{split}
\end{equation}
Here $G_0 = \mathcal{A}'[0]$, the linear slope of the amplifier's nonlinear response around zero.

In sum, the loop will reach a steady-state with a large circulating classical field with amplitude $\alpha_\t{ss}$ 
and a linearized gain of
\begin{equation}
  G \equiv \lim_{t \to +\infty} \frac{\alpha_{\text{out}}^- (t)}{\alpha_\text{in} (t)} = \frac{\mathcal{A} [\alpha_{ss}]}{\alpha_{ss}} = \frac{1}{\sqrt{\eta}}.
\end{equation}

We emphasize two key points of this classical analysis of saturating feedback oscillators. First, this system will amplify any small initial fluctuations and kick itself away from its initial, unstable, zero-field state. It will eventually reach the stable equilibrium point with a large-amplitude output field, $\alpha_{ss}$. This large output field carries the system's oscillation and is the fundamental feature of a positive feedback oscillator. Second, at the large-amplitude equilibrium point, the gain medium is linear for small perturbations, with a gain given by the requirement for the system to conserve round-trip power in steady-state i.e. $G = 1/\sqrt{\eta}$. The fact that any amplifier satisfying properties (1) through (4) embedded in a feedback loop saturates to a point where it can be treated as having a linear gain allows us to proceed with a linear quantum mechanical analysis of the oscillator's phase and amplitude fluctuations.

\subsection{Linear response}\label{sec:phasePres}

Consider the positive feedback amplifier configuration shown in \Cref{fig:fbamp}A. The output of a phase-insensitive amplifier with gain $G$ is coupled back into its input after attenuation by a factor $\sqrt{\eta}$ and a delay of $\tau$. The remaining fraction of the signal is coupled out of the loop to derive the out-of-loop field $a_\t{out}$.

The equations of motion for the system are obtained by going around the loop in \Cref{fig:fbamp}A. For the Heisenberg-picture operators in the time domain, we have
\begin{equation}\label{amp:eqsTime}
\begin{split}
	\hat{a}_\text{out}^- (t) &= G \, \hat{a}_\text{in}(t) + \sqrt{G^2 - 1} \, \hat{a}_\text{G}^\dagger (t) \\
	\hat{a}_\text{out}^+ (t) &= -\sqrt{\eta}\, \hat{a}_\text{out}^- (t) + \sqrt{1-\eta}\, \hat{a}_0 (t) \\
	\hat{a}_\text{out}(t) &= \sqrt{1-\eta}\, \hat{a}_\text{out}^- (t) + \sqrt{\eta}\, \hat{a}_0(t) \\
	\hat{a}_\text{in} (t) &= \hat{a}_\text{out}^+ (t - \tau)
\end{split}
\end{equation}
The ancillary mode $\hat{a}_\t{G}$ describes the unavoidable noise associated with any phase-insensitive linear amplification process \cite{Hau62,Cav82,Clerk10}.

\Cref{amp:eqsTime} can be solved in the frequency domain for the output field $\hat{a}_\t{out} [\Omega]$ in terms 
of the inputs $\hat{a}_0 [\Omega]$ and $\hat{a}_\t{G} [\Omega]$. The result is
\begin{equation}
\begin{split}
	\hat{a}_\text{out} [\Omega] =& \frac{\sqrt{\eta} + G e^{i \Omega \tau}}{1 + G \sqrt{\eta} e^{i \Omega \tau}} \hat{a}_0 [\Omega] + \frac{\sqrt{G^2 - 1} \sqrt{1 - \eta}}{1 + G \sqrt{\eta} e^{i \Omega \tau}} \hat{a}_\text{G}^\dagger [\Omega] \\
	&\equiv H_0 [\Omega] \hat{a}_0 [\Omega] + H_\text{G} [\Omega] \hat{a}_\text{G}^\dagger [\Omega],
\end{split}
\end{equation}
where $H_{0,\t{G}}$ are the corresponding linear response transfer functions.
In steady state, $G = 1/\sqrt{\eta}$ as discussed above, and the feedback path is characterized by two quantities, the beam-splitter transmissivity $\eta$ and the delay $\tau$, so
\begin{equation}\label{H0HgTfs}
\begin{split}
	H_0 [\Omega] = \frac{\sqrt{\eta} + \frac{1}{\sqrt{\eta}} e^{i \Omega \tau} }{1 +  e^{i \Omega \tau}},\quad 
	H_\text{G} [\Omega] = \frac{1/\sqrt{\eta} - \sqrt{\eta}}{1 + e^{i \Omega \tau}}.
\end{split}
\end{equation}

\section{Covariance Matrices of Squeezed States}\label{app:sqz}

\subsection{Covariance Matrices}

This section defines conventions for single and two mode squeezing operators and presents expressions for the covariance matrices of squeezed and entangled states, which encode these states' quadrature spectra.

We consider the spectral covariance of modes $\{\hat{a}_i\}$ in terms of their quadratures. 
Defining $\hat{Q}_i \eqdef (\hat{q}_i, \hat{p}_i)$, the the spectral covariance matrix $\mathbf{V}[\Omega]$ is defined by its 
elements \cite{Simon94,Braunstein05} $V_{ij}[\Omega] \eqdef \bar{S}_{Q_i Q_j}[\Omega]$.  

In our case, the modes are $\{\hat{a}_0,\hat{a}_\t{G}\}$.
When they are in vacuum states, $\ket{\t{vac}} = |0 \rangle_0 | 0 \rangle_\t{G}$ \cite{Braunstein05}
\begin{equation}
  \mathbf{V}_\t{vac}[\Omega] = \frac{1}{2} \t{diag}(1,1,1,1).
\end{equation}
We are also interested in the state 
\begin{equation}
	\ket{\psi} \eqdef \hat{S}_\text{E}(r_\text{E})\hat{S}_\text{G}(r_\text{G})\hat{S}_0(r_0)\ket{\text{vac}},
\end{equation}
where
\begin{equation}
\begin{split}
  \hat{S}_0(r_0) &= e^{-\frac{r_0}{2} \left(\hat{a}_0^2 - \hat{a}^{\dagger 2}_0 \right)} \\
  \hat{S}_\text{G}(r_\text{G}) &= e^{-\frac{r_\text{G}}{2} 
  	\left(\hat{a}_\text{G}^2 - \hat{a}^{\dagger 2}_\text{G} \right)} \\
  \hat{S}_\text{E} (r_\text{E}) &= e^{\frac{r_\text{E}}{2} \left(\hat{a}_0 \hat{a}_\text{G} 
  	- \hat{a}_0^{\dagger} \hat{a}_\text{G}^\dagger \right)}.
\end{split}
\end{equation}
Here, $\hat{S}_{0,\t{G}}$ are single-mode squeeze operators, 
whose sign conventions are chosen for convenience so that for a feedback oscillator, 
output phase fluctuations are suppressed when $r_0 > 0$ or $r_\t{G} > 0$. 
$\hat{S}_\t{E}$ represents two mode squeezing (or EPR entanglement),
and output phase fluctuations are suppressed when $r_\t{E}>0$.
The covariance matrix of $\ket{\psi}$ is \cite{Braunstein05}
\begin{equation}\label{eq:Vsqzcorr}
\begin{split}
  \mathbf{V} [\Omega]
  & = \begin{bmatrix} \mathbf{I}(2r_\text{G},2r_0,\frac{r_\text{E}}{2}) & \mathbf{Z}(2 r_0, 2 r_\text{G}, r_\text{E}) \\
  \mathbf{Z}(2 r_0, 2 r_\text{G}, r_\text{E})^T & \mathbf{I}(2r_0,2r_\text{G},\frac{r_\text{E}}{2}) \end{bmatrix},
\end{split}
\end{equation}
where we have defined the matrices $\mathbf{I}(x,y,z)$ and $\mathbf{Z}(x, y, z)$ by
\begin{equation}
\begin{split}
  &\mathbf{I}(x,y,z) = \frac{1}{2} \times \\
  & \, \begin{bmatrix} e^{x}\sinh^2 (z)+ e^{y}\cosh^2(z) & 0 \\ 0 & e^{-x}\sinh^2 (z)+ e^{-y}\cosh^2(z) \end{bmatrix} \\
  &\mathbf{Z}(x,y,z) = \frac{1}{4} \times \\
  &\, \begin{bmatrix} \left( e^x + e^y \right) \sinh (z) & 0 \\ 0 & - \left(e^{-x} + e^{-y} \right) \sinh(z) \end{bmatrix}.
\end{split}
\end{equation}
The quadrature spectra for the two mode state with squeezed and EPR correlated modes parametrized by $r_0$, $r_\t{G}$ and $r_\t{E}$ can be read off of \cref{eq:Vsqzcorr}.

\subsection{Uncertainty Products for Entangled States}

\begin{figure}[t!]
  \centering
  \includegraphics[width=0.85\columnwidth]{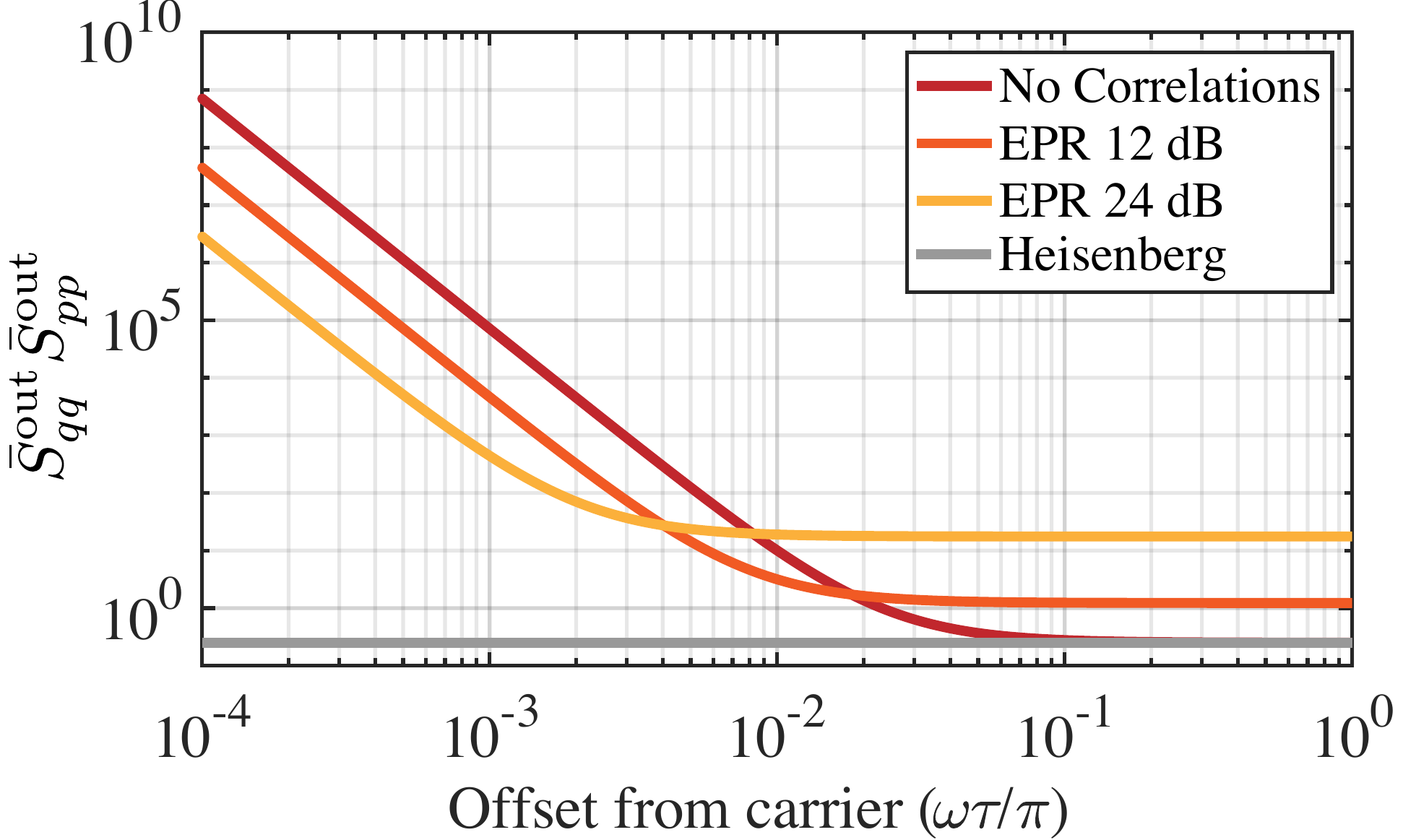}
  \caption{\label{fig:epr} Uncertainty product $\bar{S}_{qq}^\t{out} [\Omega] \bar{S}_{pp}^\t{out} [\Omega]$ for
  a feedback oscillator with phase-sensitive amplifier fed with various levels of EPR entangled states for the 
  in-coupled and ancillary modes. As the level of squeezing used to generate entanglement is increased from zero 
  (red) to 12 dB (orange) and then to 24dB (yellow), the uncertainty product decreases near the carrier and 
  increases far from it. In all cases, the uncertainty product is above the Heisenberg limit (grey).
  Note that the uncertainty product is symmetric about the carrier.}
\end{figure}

As mentioned in the discussion of \cref{eq:sqzCorrSpec}, our near-resonant approximation makes it appear as if it is possible to violate Heisenberg uncertainty for sufficiently large levels of EPR entanglement between the in-coupled and ancillary modes. We now show that this is not the case.

Dropping the near-resonant approximation, the quadrature spectra of the output mode are related to those of the in-coupled and ancillary modes by
\begin{equation}
\begin{split}
  \Bar{S}_{qq}^\text{out} [\Omega] =& \abs{H_0 [\Omega]}^2 \Bar{S}_{qq}^0 [\Omega] + \abs{H_\text{G} [\Omega]}^2 \Bar{S}_{qq}^\text{G} [\Omega] \\
                                    &+ 2 \, \text{Re} \left[H_0 [\Omega] H_\text{G} [\Omega]^* \bar{S}_{qq}^{0,\text{G}} [\Omega] \right] \\
  \Bar{S}_{pp}^\text{out} [\Omega] =& \abs{H_0 [\Omega]}^2 \Bar{S}_{pp}^0 [\Omega] + \abs{H_\text{G} [\Omega]}^2 \Bar{S}_{pp}^\text{G} [\Omega] \\
                                    &- 2 \, \text{Re} \left[H_0 [\Omega] H_\text{G} [\Omega]^* \bar{S}_{pp}^{0,\text{G}}[\Omega] \right],
\end{split}
\end{equation}
Using the spectra from \cref{eq:Vsqzcorr} with $r_0 = r_\t{G} = 0$, we find
\begin{equation}
\begin{split}
  \Bar{S}_{qq}^\text{out} [\Omega] =& \frac{e^{r_\text{E}}}{4 \eta} \\
                                    &+ \frac{e^{-r_\text{E}}}{4} \left[\left(\frac{1}{\eta} - 2 \right) \tan^2 \left( \frac{\Omega \tau}{2} \right) + \eta \sec^2 \left( \frac{\Omega \tau}{2} \right) \right]
\end{split}
\end{equation}
and $\Bar{S}_{pp}^\text{out} [\Omega] = \Bar{S}_{qq}^\text{out} [\Omega]$. Using this expression, we find that the product $\Bar{S}_{qq}^\text{out} [\Omega] \Bar{S}_{pp}^\text{out} [\Omega]$ is minimized for $\Omega \tau = 2 \pi n$ for $n \in \mathbb{Z}$. The minimum value of this product is
\begin{equation}
  \min_\Omega \left[\Bar{S}_{qq}^\text{out} [\Omega] \Bar{S}_{pp}^\text{out} [\Omega] \right] = \left( \frac{e^{r_\text{E}}}{4 \eta} + \frac{\eta e^{-r_\text{E}}}{4} \right)^2
\end{equation}
This product is itself minimized when $r_\text{E} = \ln(\eta)$, with the result
\begin{equation}
  \min_{\{\Omega,\eta,r_\text{E}\}} \left[\Bar{S}_{qq}^\text{out} [\Omega] \Bar{S}_{pp}^\text{out} [\Omega] \right] = \frac{1}{4}.
\end{equation}
Thus, for general values of $\Omega$, $\eta$, and $r_\t{E}$, we have the relation
\begin{equation}
  \Bar{S}_{qq}^\text{out} [\Omega] \Bar{S}_{pp}^\text{out} [\Omega] \geq \frac{1}{4},
\end{equation}
which is exactly the Heisenberg uncertainty bound. The fact that EPR entangling the in-coupled and ancillary modes does not allow the feedback oscillator to violate Heisenberg uncertainty is shown for several specific levels of squeezing in \cref{fig:epr}.

In \cref{sec:sqzEpr}, we have seen that EPR entanglement can suppress the amplitude and phase quadratures simultaneously and to arbitrary levels in the near-resonant approximation. However, we have now shown that when we consider the full, unapproximated system dynamics, the system always obeys the Heisenberg uncertainty bound for any amount of EPR entanglement. Physically, this means that our system does not violate one of the key tenets of quantum theory, which is an important check on the validity of our results.

\section{Details of feedback oscillator based on phase-sensitive amplifier}\label{app:fbOscPhSensAmps}

This section provides extant details of the quantum noise properties of feedback oscillators 
with a phase-sensitive amplifier embedded in the loop. 

\subsection{Decomposing a Phase-Sensitive Amplifier as a Phase-Insensitive Amplifier and a Squeezer}
\label{sec:phaseSensitiveAmpDecomp}

In this section, we show that a quantum-limited phase-sensitive amplifier can always be decomposed into an 
ideal phase-sensitive amplifier followed by an ideal single-mode squeezer. 

\begin{figure}[t!]
  \centering
  \includegraphics[width=0.8\columnwidth]{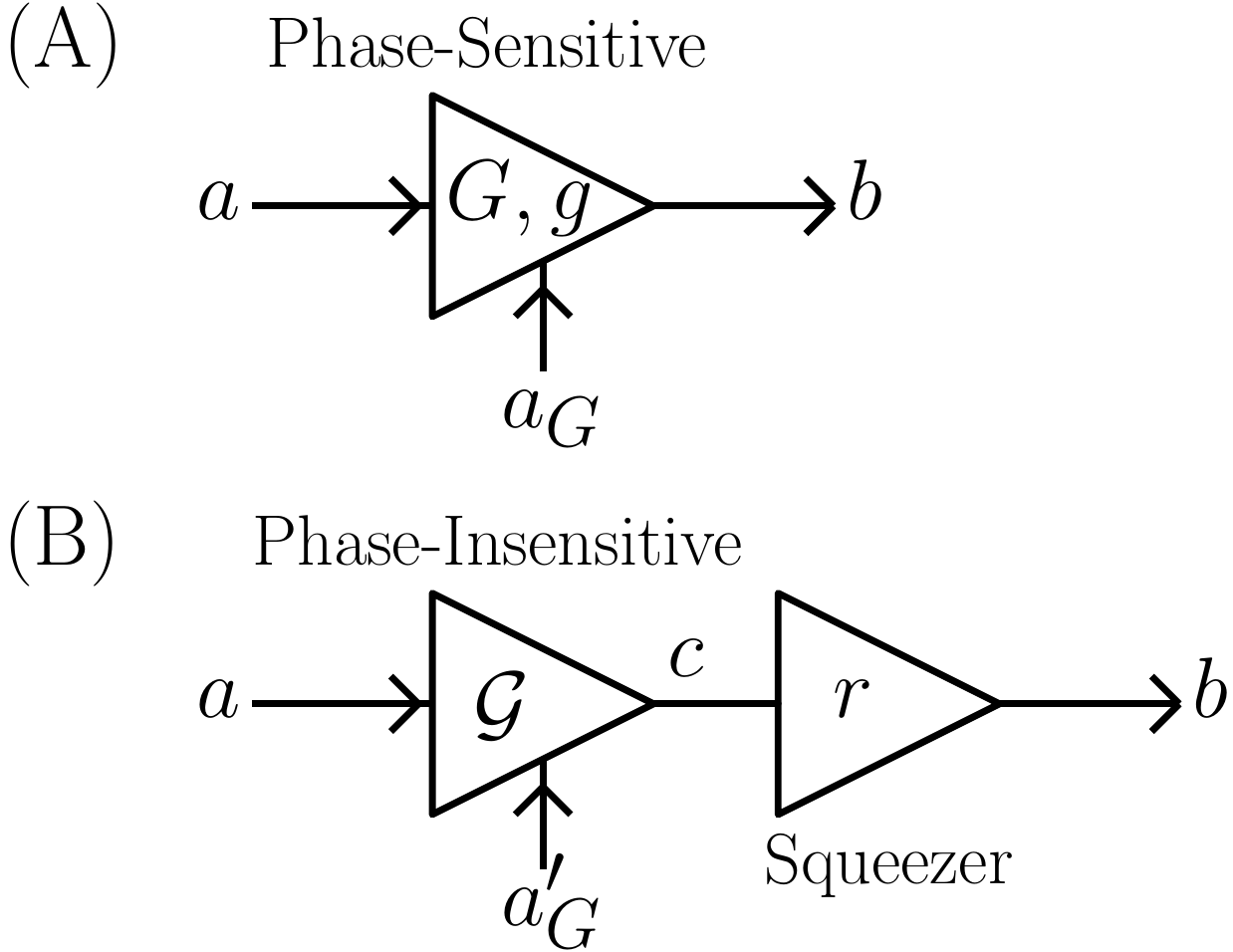}
  \caption{\label{fig:ampDecomp} (A) A phase-sensitive amplifier. (B) A decomposition of the phase-sensitive amplifier from (A) into a phase-insensitive amplifier followed by a squeezer.}
\end{figure}

Consider a phase-sensitive amplifier that amplifies the input $\hat{a}$ to produce the output 
$\hat{b} = G \hat{a} + g \hat{a}^\dagger + \hat{a}_\t{G}^\dagger$, where $\hat{a}_\t{G}$ is the 
amplifier's ancillary mode (see \cref{fig:ampDecomp}A). 
(Note that the ancillary mode, $\hat{a}_\t{G}$, does not necessarily have bosonic statistics here.)
We assert that such a phase-sensitive amplifier is equivalent to an ideal phase-insensitive 
amplifier with gain $\mathcal{G} = \sqrt{G^2-g^2}$ that sends mode $\hat{a}$ to the mode $\hat{c} = \mathcal{G} \hat{a} + \hat{a}_\t{G}^{\prime \dagger}$, followed by an ideal squeezer which sends the mode $\hat{c}$ to the output mode $\hat{b} = \cosh (r) \hat{c} +  \sinh(r) \hat{c}^\dagger$ with $\tanh(r) = g/G$ (see \cref{fig:ampDecomp}B). 
For this equivalency to hold, we require $\hat{a}_\t{G}^\dagger = (G \hat{a}_\t{G}^{\prime \dagger} + g \hat{a}_\t{G}^\prime)/\mathcal{G}$.

We can prove this by explicit computation. In the case of the phase-sensitive amplifier, the output mode $\hat{b}$ is given in terms of the input mode $\hat{a}$ by $\hat{b} = G \hat{a} + g \hat{a}^\dagger + \hat{a}_\t{G}^\dagger$. In the case of a phase-insensitive amplifier followed by a squeezer, we have (assuming $G$ and $g$ are positive)
\begin{equation}
\begin{split}
	\hat{b} &= \cosh(r) \hat{c} + \sinh(r) \hat{c}^\dagger \\
		&= \cosh(r) (\mathcal{G} \hat{a} + \hat{a}_\text{G}^{\prime \dagger}) + \sinh(r) (\mathcal{G} \hat{a}^\dagger + \hat{a}_\text{G}^\prime) \\
		&= G \hat{a} + g \hat{a}^\dagger + \frac{G \hat{a}_\text{G}^{\prime \dagger} + g \hat{a}_\text{G}^\prime}{\sqrt{G^2-g^2}} \\
		&= G \hat{a} + g \hat{a}^\dagger + \hat{a}_\text{G}^\dagger,
\end{split}
\end{equation}
Additionally, we can show that $\hat{a}_\t{G}$ and $\hat{a}_\t{G}^\prime$ have the same commutation relations.
\begin{equation}
\begin{split}
  \left[ a_\text{G},a_\text{G}^\dagger \right] &= \frac{\left[G a_\text{G}^\prime + g a_\text{G}^{\prime \dagger}, G a_\text{G}^{\prime \dagger} + g a_\text{G}^{\prime} \right]}{G^2-g^2} \\
  &= \frac{G^2 \left[a_\text{G}^\prime,a_\text{G}^{\prime \dagger} \right] + g^2 \left[ a_\text{G}^{\prime \dagger}, a_\text{G}^\prime \right]}{G^2 - g^2} \\
  &= \left[a_\text{G}^\prime,a_\text{G}^{\prime \dagger} \right],
\end{split}
\end{equation}
so we see that this is the correct decomposition of a phase-sensitive amplifier.

The primary benefit of this decomposition is that it clarifies 
the extent to which a phase-sensitive amplifier must add noise: only the phase-insensitive component in its decomposition
adds noise, while the squeezer is noiseless.

\subsection{Response of an oscillator with phase-sensitive amplifier}\label{app:phSensFbOsc}

For an oscillator composed of an in-loop phase-sensitive amplifier (\cref{fig:fbamp}B), 
it is more natural to study its linear response in terms of the quadratures of the various fields involved.
We define the generalized quadratures $\hat{q}_\theta$ and $\hat{p}_\theta$ by
\begin{equation}\label{eq:genQuadratutes}
  \hat{q}_\theta = \frac{\hat{a}^\dagger e^{i \theta} + \hat{a}e^{-i \theta}}{\sqrt{2}}, \quad
  \hat{p}_\theta = \hat{q}_{\theta + \pi/2},
\end{equation}
which are canonically conjugate and related to the usual amplitude/phase quadratures as $\hat{q} = \hat{q}_{\theta = 0}$ and 
$\hat{p} = \hat{p}_{\theta = 0}$.

The linear response equations \cref{amp:eqsFtPhaseSensitive} can be expressed in terms of the quadratures.
Choosing the phase angles in \cref{eq:genQuadratutes} to be $\theta_\t{out} = \theta_\t{out}^+ = \theta_\t{out}^- = \theta_\t{s} = \theta_\t{G} = \theta_\t{in} = \theta_0 = \phi_\t{s}/2$, the amplitude quadrature operators describing the coupled modes of a phase-sensitive oscillator satisfy
\begin{equation}\label{eq:phaseSensitiveLoopQquad}
\begin{split}
	\hat{q}_{\theta,\text{out}}^- [\Omega] &= e^{r_\text{s}} \hat{q}_{\theta,\text{s}} [\Omega] \\
	\hat{q}_{\theta,\text{s}} [\Omega] &= G \, \hat{q}_{\theta,\text{in}}[\Omega] + \sqrt{G^2 - 1}\, \hat{q}_{\theta,\text{G}}[\Omega] \\
	\hat{q}_{\theta,\text{out}}^+ [\Omega] &= -\sqrt{\eta}\, \hat{q}_{\theta,\text{out}}^- [\Omega] + \sqrt{1-\eta}\, \hat{q}_{\theta,0} [\Omega] \\
	\hat{q}_{\theta,\text{out}}[\Omega] &= \sqrt{1-\eta}\, \hat{q}_{\theta,\text{out}}^- [\Omega] + \sqrt{\eta}\, \hat{q}_{\theta,0}[\Omega] \\
	\hat{q}_{\theta,\text{in}} [\Omega] &= e^{i\Omega \tau} \hat{q}_{\theta,\text{out}}^+ [\Omega],
\end{split}
\end{equation}
and the phase quadrature operators for this oscillator satisfy
\begin{equation}\label{eq:phaseSensitiveLoopPquad}
\begin{split}
	\hat{p}_{\theta,\text{out}}^- [\Omega] &= e^{-r_\text{s}} \hat{p}_{\theta,\text{s}} [\Omega] \\
	\hat{p}_{\theta,\text{s}} [\Omega] &= G \, \hat{p}_{\theta,\text{in}}[\Omega] - \sqrt{G^2 -1}\, \hat{p}_{\theta,\text{G}}[\Omega] \\
	\hat{p}_{\theta,\text{out}}^+ [\Omega] &= -\sqrt{\eta}\, \hat{p}_{\theta,\text{out}}^- [\Omega] + \sqrt{1-\eta}\, \hat{p}_{\theta,0} [\Omega] \\
	\hat{p}_{\theta,\text{out}}[\Omega] &= \sqrt{1-\eta}\, \hat{p}_{\theta,\text{out}}^- [\Omega] + \sqrt{\eta}\, \hat{p}_{\theta,0}[\Omega] \\
	\hat{p}_{\theta,\text{in}} [\Omega] &= e^{i\Omega \tau} \hat{p}_{\theta,\text{out}}^+ [\Omega].
\end{split}
\end{equation}
We note that unlike the phase-insensitive amplifier case where all phase angles were relative to the oscillator's output phase $\theta_\t{out}$, the phase angles are now all defined relative to the squeeze angle $\phi_\t{s}$. Having made this point, we will now take $\phi_\t{s} = 0$.

We now define the $\hat{q}$ and $\hat{p}$ quadrature transfer functions $H_0^q [\Omega]$, $H_\t{G}^q [\Omega]$, $H_0^p [\Omega]$, and $H_\t{G}^p [\Omega]$ by
\begin{equation}\label{eq:defQuadTfs}
\begin{split}
	\hat{q}_\text{out} [\Omega] &= H_0^q [\Omega] \hat{q}_0 [\Omega] + H_\text{G}^q [\Omega] \hat{q}_\text{G} [\Omega] \\
	\hat{p}_\text{out} [\Omega] &= H_0^p [\Omega] \hat{p}_0 [\Omega] - H_\text{G}^p [\Omega] \hat{p}_\text{G} [\Omega].
\end{split}
\end{equation}
The sign convention is chosen such that $H_0^q = H_0^p = H_0$ and $H_\text{G}^q = H_\text{G}^p = H_\text{G}$ in the case that $r_\t{s} \rightarrow 0$ and the phase-sensitive amplifier reduces to a phase-insensitive amplifier.

With a phase-sensitive amplifier in the feedback loop, the saturation condition is modified. Still modeling the phase-sensitive amplifier as a phase-insensitive amplifier and a squeezer, the phase-insensitive amplifier will saturate when the loop gain for the amplified quadrature is equal to unity. Mathematically, this condition is
\begin{equation}\label{eq:phaseSensSaturationCondition}
	G e^{r_\text{s}} \sqrt{\eta} = 1,
\end{equation}
which tells us that in steady state, we have $G = e^{-r_\t{s}}/\sqrt{\eta}$. If the amplifier 
is purely phase-sensitive, and thus represented by an ideal squeezer, then $G = 1$, and $e^{r_\t{s}} = 1/\sqrt{\eta}$. 
We denote this value of $r_\t{s}$ by $r_\t{max}$. 
For squeezing above this value, saturation effects will reduce $r_\t{s}$ back to $r_\t{max}$. 
Explicitly, $r_\t{max}$ is given by
\begin{equation}
	r_\text{max} \equiv - \frac{1}{2} \ln (\eta).
\end{equation}

Solving \cref{eq:phaseSensitiveLoopQquad,eq:phaseSensitiveLoopPquad} and using the condition of \cref{eq:phaseSensSaturationCondition} to eliminate $G$ from these equations, we find that the quadrature transfer functions are given by
\begin{equation}\label{eq:PhQuadTf}
\begin{split}
	H_0^q [\Omega] &= \frac{e^{i \Omega \tau} / \sqrt{\eta} + \sqrt{\eta}}{1 + e^{i \Omega \tau}} = H_0 [\Omega] \\
	H_\text{G}^q [\Omega] &= \frac{\sqrt{1/\eta - e^{2 r_\text{s}}} \, \sqrt{1 - \eta}}{1 + e^{i \Omega \tau}} = \sqrt{\frac{1-\eta e^{2 r_\text{s}}}{1 - \eta}} H_\text{G} [\Omega] \\
	H_0^p [\Omega] &= \frac{e^{i \Omega \tau} / \sqrt{\eta} + e^{2 r_\text{s}} \sqrt{\eta}}{e^{2 r_\text{s}} + e^{i \Omega \tau}} \\
	H_\text{G}^p [\Omega] &= \frac{\sqrt{1/\eta - e^{2 r_\text{s}}} \, \sqrt{1 - \eta}}{e^{2 r_\text{s}} + e^{i \Omega \tau}}.
\end{split}
\end{equation}
From these equations, we see that $H_\text{G}^q [\Omega] \rightarrow 0$ and $H_\text{G}^p [\Omega] \rightarrow 0$ as $r \rightarrow r_\t{max}$. Physically, the amplifier does not add any noise to the output as it becomes completely phase-sensitive.

We see that near resonance where $\Omega \tau \approx (2 n + 1) \pi$, $H_0^q$ and $H_\t{G}^q$ still scale as $1/\omega$, whereas $H_0^p$ and $H_\t{G}^p$ have different denominators which do not approach zero as $\omega$ becomes small. Near resonance, where $\omega \tau \ll 1$, the quadrature transfer functions are given to leading order in $\omega \tau$ by
\begin{equation}\label{eq:approxPhQuadTf}
\begin{split}
	H_0^q [\omega] &\approx \frac{1/\sqrt{\eta} - \sqrt{\eta}}{i \omega \tau} \\
	H_\text{G}^q [\Omega] &\approx -\frac{\sqrt{1/\eta - e^{2 r_\text{s}}} \, \sqrt{1-\eta}}{i \omega \tau} \\
	H_0^p [\omega] &\approx \frac{-\sqrt{\eta}(1/\eta - e^{2 r_\text{s}}) - i \omega \tau/\sqrt{\eta}}{e^{2 r_\text{s}} - 1 - i \omega \tau} \\
	H_\text{G}^p [\omega] &\approx \frac{\sqrt{1/\eta - e^{2 r_\text{s}}} \, \sqrt{1 - \eta}}{e^{2 r_\text{s}} - 1 - i \omega \tau}.
\end{split}
\end{equation}

The spectra of the output amplitude and phase are related to those of the in-coupled and ancillary modes as:
\begin{equation}\label{eq:ampSpectralRelationsPhaseSensitiveAmp}
\begin{split}
	\bar{S}_{qq}^\text{out} [\Omega] =& |H_0^q [\Omega]|^2 \bar{S}_{qq}^0 [\Omega] + |H_\text{G}^q [\Omega]|^2 \bar{S}_{qq}^\text{G} [\Omega] \\
	&+ 2 \text{Re} \left[H_0^q [\Omega] H_\text{G}^q [\Omega]^* \bar{S}_{qq}^{0,G} [\Omega] \right],
\end{split}
\end{equation}
\begin{equation}\label{eq:phSpectralRelationsPhaseSensitiveAmp}
\begin{split}
	\bar{S}_{pp}^\text{out} [\Omega] =& |H_0^p [\Omega]|^2 \bar{S}_{pp}^0 [\Omega] + |H_\text{G}^p [\Omega]|^2 \bar{S}_{pp}^\text{G} [\Omega] \\
	&- 2 \text{Re} \left[H_0^p [\Omega] H_\text{G}^p [\Omega]^* \bar{S}_{pp}^{0,G} [\Omega] \right].
\end{split}
\end{equation} 
Near resonance, the output amplitude quadrature spectrum of the phase-sensitive feedback oscillator is given by
\begin{equation}
\begin{split}
  \bar{S}_{qq}^\text{out} [\Omega] = |H_0 [\Omega]|^2 \Bigg( \bar{S}_{qq}^0 [\Omega] + \left( \frac{1 - \eta e^{2 r_\text{s}}}{1-\eta} \right) \bar{S}_{qq}^\text{G} [\Omega]& \\
  - 2 \sqrt{\frac{1-\eta e^{2 r_\text{s}}}{1-\eta}} \text{Re} \left[\bar{S}_{qq}^{0,G} [\Omega] \right] &\Bigg).
\end{split}
\end{equation}
We see that the output amplitude quadrature of the phase-sensitive oscillator retains the same spectral shape as in the phase-insensitive oscillator, determined by $|H[\Omega]|$. However, as the oscillator becomes more phase-sensitive, its amplifier contributes less noise to the output amplitude quadrature.

Further, from \cref{eq:approxPhQuadTf}, we see that the output phase-quadrature spectrum of the phase-sensitive oscillator has a fundamentally different spectral shape than it does for a phase-insensitive oscillator. The phase-quadrature transfer functions no longer have poles at $\omega\tau = 0$ and in the case of a perfectly phase-sensitive feedback oscillator, the phase-quadrature transfer function for the in-coupled mode, $H_0^p$ has a zero at $\omega\tau = 0$.

Using the fact that $\hat{a}_0$ and $\hat{a}_\t{out}$ are freely propagating bosonic modes and the transfer functions from \cref{eq:PhQuadTf}, we can compute the statistics of the amplifier's ancillary mode. As in the case of the phase-insensitive amplifier, $\hat{a}_\t{G}$ is not freely propagating, so its statistics need not be bosonic. Computing the ancillary mode's statistics, we find
\begin{equation}
	[\hat{a}_\text{G} [\Omega], \hat{a}_\text{G}^\dagger [\Omega^\prime]] =  2\pi \cdot \delta [\Omega + \Omega^\prime],
\end{equation}
so as for a phase-insensitive feedback oscillator, the amplifier's ancillary mode obeys bosonic statistics despite the fact that it is an in-loop field.

\subsection{Quadrature spectra for uncorrelated in-coupled and ancillary modes}\label{sec:quadSpecPhSens}

In the absence of correlations between the in-coupled and ancillary modes, \cref{eq:ampSpectralRelationsPhaseSensitiveAmp,eq:phSpectralRelationsPhaseSensitiveAmp} reduce to
\begin{equation}
\begin{split}
	\bar{S}_{qq}^\text{out} [\Omega] &= |H_0^q [\Omega]|^2 \bar{S}_{qq}^0 [\Omega] + |H_\text{G}^q [\Omega]|^2 \bar{S}_{qq}^\text{G} [\Omega] \\
	\bar{S}_{pp}^\text{out} [\Omega] &= |H_0^p [\Omega]|^2 \bar{S}_{pp}^0 [\Omega] + |H_\text{G}^p [\Omega]|^2 \bar{S}_{pp}^\text{G} [\Omega].
\end{split}
\end{equation}
So the amplitude and phase quadrature spectra of the phase-sensitive feedback oscillator with vacuum state in-coupled and ancillary modes are given by
\begin{equation}
\begin{split}
	\bar{S}_{qq}^\text{out} [\omega] =& \frac{1}{2}|H_0^q [\omega]|^2 + \frac{1}{2} |H_\text{G}^q [\omega]|^2 \\
	\bar{S}_{pp}^\text{out} [\omega] =& \frac{1}{2}|H_0^p [\omega]|^2 + \frac{1}{2}  |H_\text{G}^p [\omega]|^2,
\end{split}
\end{equation}
near resonance.

Plugging in the explicit expressions for the quadrature transfer functions near resonance from \cref{eq:approxPhQuadTf}, the output amplitude and phase quadrature spectra are given by
\begin{equation}\label{eq:SppPhaseSensNearRes}
\begin{split}
	\bar{S}_{qq}^\text{out} [\Omega] \approx& \frac{(1-\eta)^2 + (1 - \eta e^{2 r_\text{s}})(1 - \eta)}{2 \eta \tau^2 \, \omega^2} \\
	\bar{S}_{pp}^\text{out} [\Omega] \approx& \frac{ (1 - \eta e^{2 r_\text{s}})^2 + \tau^2 \, \omega^2 + (1 - \eta e^{2 r_\text{s}})(1 - \eta)}{2 \eta \left[ (e^{2 r_\text{s}} - 1)^2 + \tau^2 \, \omega^2 \right]}.
\end{split}
\end{equation}

\end{document}